# A Cosmological Bragg Law: Interpreting the CMB as a Diffractogram of Foliated Spacetime

## David Izabel

**Abstract:**

Recent work by Ringermacher and Mead has revealed discrete oscillations in the cosmological scale factor, suggesting that spacetime may exhibit vibrational properties akin to those of a crystal. Building on this idea, we propose a novel geometric interpretation of the Cosmic Microwave Background (CMB) angular power spectrum by treating it as an inverted X-ray diffractogram of early spacetime. In this framework, the acoustic peaks correspond to constructive interference from a stratified, resonant structure of the universe at the time of recombination. We formulate a cosmological analogue of Bragg's law, linking angular multipoles to acoustic shell spacing and comoving distance. This model predicts the positions of the first four acoustic peaks observed by the Planck satellite with remarkable accuracy. A global correction factor of $\approx 0.9$, needed to align predictions with observations, may reflect the influence of geometric torsion as described in Einstein–Cartan theory. While not a replacement for full radiative transfer modeling, this approach offers a complementary geometric framework for understanding the harmonic structure of the CMB and the possible crystalline nature of primordial spacetime.

## 1. Introduction

General relativity **[1] [2]** has now been well demonstrated and verified experimentally for more than 100 years. Space-time is deformed, that's indisputable. This can be seen with the Lense-Thiring effect **[3]** tested and verified with the Gravity probB experiment **[4]**, the gravitational waves predicted by Einstein **[5]** and measured by the LIGO interferometers GW150914 **[6]** and GW170817 **[7]**, etc. But these deformations are extremely tenuous, for example they are of the order of $10^{-21}$ **[6]** in the case of gravitational waves. Moreover, these deformations disappear when the mass densities of energy that create them disappear, as shown for example by the examination of the angular deviations of stars located behind the sun during an eclipse as shown in 1919 by Eddington and his team **[8]**. In short, space-time behaves by analogy like an elastic medium, as A. Sakharov wrote as a precursor **[9]**. The theory of general relativity could therefore be seen as a theory of elasticity **[10]** or Hooke's law in 4 dimensions **[11]**. As a result, many authors have shown that indeed general relativity can be seen in a weak field by analogy as a homogeneous and isotropic deformable elastic medium characterized by its Young's modulus **[12]** to **[21]**, its Poisson's ratio **[12]** to **[15]**, its creep coefficient **[22]**, its coefficient of thermal expansion **[23]** etc. However, by analyzing deformations more closely, particularly in the case of gravitational waves, some authors **[24]** have shown that space-time actually appears to be laminated as made of independent sheets. These very thin sheets, with a Planck thickness **[12]** to **[15]**, make it possible to model the functioning of space-time today **[12]** to **[15]** **[24]**, particularly in weak gravitational fields for phenomena such as gravitational waves. These sheets or planes transverse to the direction of propagation of gravitational waves are not, however, connected to each other in the case of classical general relativity **[24]**. How do we know? in fact, Tenev and Horstemeyer **[12]** **[13]** in particular have shown that there is a tensor bridge between the perturbation tensor of the metric $h_{\mu\nu}$ and the deformation tensor of the equivalent elastic medium $\varepsilon_{\mu\nu}$: $h_{\mu\nu} = 2\varepsilon_{\mu\nu}$. Consequently, the polarizations $A^+$ and $A^\times$ that appear in planes transverse to the direction of propagation of gravitational waves in classical general relativity correspond to deformations in these same transverse planes only **[24]**. By reciprocity, since there is no polarization in the direction of propagation of gravitational waves, there are no distortions between these planes either. The typology of space, in the case of gravitational waves in classical general relativity, is therefore laminated. In the context of the analogy of an elastic medium, this poses a problem. To reconstruct a minimum 3-dimensional environment, it is necessary to restore cohesion between these sheets. This is possible by going through a modified general relativity of the Einstein-Cartan type with geometric torsion added to the Riemann tensor whose elasticity mirror is the theory of defects **[25]** to **[27]**. With such a theory, space-time presents polarizations complementary to those described above **[24]** **[28]**. What is interesting is that these complementary polarizations are this time in the direction of propagation of gravitational waves. These polarizations can be transformed into equivalent deformations according to reconstituting a 3-dimensional medium according to for example **[29]**. In this case, the mirror of space-time, by analogy with an elastic medium, is a transverse isotropic crystal **[30]** to **[32]**, consisting of sheets connected to each other and presenting locally defects **[25]** to **[27]**. The laws of passage between these polarizations and the deformations are more complex because there is local plasticization to consider. This has been studied by some authors such as **[28]**. These defects are related to the plasticization of the crystal **[24]**, **[25]** to **[27]**. This theory of defects is the exact mathematical and physical reflection of the Einstein-Cartan theory of general relativity with geometric torsion **[24]**, **[25]** to **[27]**. The structure of the Burgers vector is tensorially close to the notion of geometric torsion in Einstein Cartan's theory **[25]** to **[27]**. The equation of general relativity is then replaced by two equations **[25]** to **[27]**. Einstein's is complemented by another in spin. We can show that by analogy general relativity is a kind of Timoshenko strength of materials in 4D. Thus, as in the strength of materials when studying a structure, there are two equations to consider, one in force and another in rotational moments. This model of space-time as a crystal has been studied by many authors **[30]** to **[32]**. This model also seems quite relevant because a recent study shows that by measurements **[32]**, the intercrystal forces of modern materials such as Graphene are of the same order of magnitude as the κ coupling constant of these classical or modified general relativities.

So, at this stage of our reflection, for the case of gravitational waves, the typology of space-time is a transverse isotropic medium, made up of thin sheets that are very inconsistent with each other, locally plasticizing and a Planck scale the structure is reminiscent of that of a crystal. This is the teaching of general relativity modified with Einstein Cartan twist and studied in **[24]**. These complementary polarizations have not yet been measured, but LISA interferometers, equipped with 3 arms, should make it possible to settle this question. In any case, the space-time foliation models developed by ADM **[33]** or Tenev and Horstemeyer **[12]** **[13]** or Izabel and al **[24]** seem quite relevant to the measured deformations (transverse planes) and potentially longitudinal (between these planes) remaining to be measured.

At this stage, and this is the purpose of this paper, we can wonder if there is not another way than gravitational waves to probe the structure and therefore the typology of space-time.

Indeed, space-time has been expanding for 13.7 billion years. If, therefore, its internal structure is by analogy like a crystal **[30]** to **[32]**, then it must have had this structure from its birth, which has since only swelled.

It turns out that we have a natural image of the universe 380,000 years after its birth. The Cosmic Microwave Background (CMB) is the oldest image of the observable universe. Its precise study, by the COBE, WMAP and especially Planck satellites for the high precision of the results **[34]** to **[37]**, made it possible to measure with unprecedented precision the temperature anisotropies and polarizations of the CMB, paving the way for precision cosmology. In the case of the cosmic microwave background, the light and therefore the photons trapped within the baryonic matter and the dark matter, traveling within the very intrinsic structure of space-time, ended up being able to exit from the inside to the outside of it, thus freezing in this radiation the structure, the anisotropies, the concentration and the lack of dark or baryonic matter. So, they mapped the structure of the original crystal of space-time.

In his publications **[72] [73]** Ringermacher and Mead's have spotted vibrations that suggest that space-time does indeed vibrate like a crystal. So, if this is the case, an analysis by light, as in the case of X-rays, should make it possible to analyze its structure.

Although the discrete oscillation model of the scaling factor proposed by Ringermacher and Mead has not yet been confirmed by other observations, it nevertheless provides an interesting heuristic motivation for exploring periodic structures in the history of cosmic expansion. Our approach does not depend on the validity of their model, but draws on it as a conceptual starting point.

At the same time, taking the problem the other way around, starting from the analogy of space-time as a crystal as described above, X-ray diffraction in crystallography has established itself as the standard method for probing the regular atomic structure of materials and crystals **[38] [39]**. Except that this time, it is by bombarding these crystals with X-ray rays from the outside to the inside of the crystal, from different angles, that from the intensity of the rays that have passed through the crystal on the one hand, spots appear on the control screen indicating the geometric position of the atoms on the other hand. By this method, it is possible to establish the structure of the crystal studied.

From there an idea emerged, at the origin of this paper. In this paper, we explore the idea that the spectrum of the CMB is conceptually comparable to an inverted X-ray diffractogram (the light coming out of the crystal), revealing not atoms, but primordial "grains" of baryonic matter distributed according to the structure of dark matter which is perhaps the very structure of space-time, this unknown crystal. Can we therefore study the power spectra in energy and polarization as X-ray diffractograms of the space-time equivalent crystal in its youth? If the analysis of these power spectra or inverted X-ray diffractograms shows peaks whose distributions and widths are similar to those of lamellae crystals, then we can be more confident about the foliated typologies of the universe and therefore about the representativeness of theories of the type ADM **[33]**, Tenev and Horstemeyer **[12] [13]**, Izabel, and al **[24]**. This is what we will study in the next paragraphs. While our model does not attempt to reproduce the full detailed shape of the CMB spectrum (which depends on Boltzmann solvers and radiative transfer), it offers a geometric interpretation of the harmonic peak positions based on large-scale structure resonance. In this context, the present work proposes a novel approach to probing the structure of spacetime by interpreting the Cosmic Microwave Background (CMB) as an inverted X-ray diffractogram of a primordial cosmological crystal. Rather than relying solely on gravitational wave observations to infer the laminated nature of spacetime, we explore whether the angular power spectrum of the CMB—particularly the positions and spacing of its acoustic peaks—can be understood as the harmonic response of a stratified medium. Drawing on analogies with crystallography, we formulate a cosmological version of Bragg's law and test its predictive power against Planck satellite data. This geometric framework offers a complementary perspective to standard cosmological models, potentially revealing the large-scale coherence and mechanical properties of early spacetime. The following sections detail the methodology, theoretical background, and implications of this analogy.

This approach is therefore based on a rigorous geometric analogy, but it is not limited to a simple metaphor. We propose a quantitative formulation (cosmological Bragg law) digitally tested on Planck data. This analogy is in line with the tradition of effective models in physics (such as Sakharov elasticity or Kleinert crystals), which aim to reveal underlying structures from observable signatures

## 2. Method

This study investigates whether the observed angular structure of the Cosmic Microwave Background (CMB) can be interpreted as the harmonic response of a foliated spacetime medium—analogous to X-ray diffraction patterns in layered crystals. To explore this hypothesis, we developed the following methodology:

1) We use Planck 2018 public data (thermal power spectra and polarization) **[34]** to **[37]** to extract:

- The first peak of the TT spectrum, corresponding to the fundamental resonance **[40]**,
- The spacing between successive peaks (harmonics) **[40]**,

2) We will then translate these results on peaks from a cosmological point of view in the light of knowledge in crystallography **[38] [39]** and in particular:

- For the distance between peaks as the "cosmological interleaf distance", a proposal of the cosmological Bragg laws
- For the width of the peaks as the size of the initial coherence domain of the cosmological crystal.

This approach allows us to interpret primordial acoustic structures as crystallographic signatures of "middle space-time" at the time of photon-baryonic matter and space-time decoupling.

3) We will then look for materials on earth and therefore crystal structures that have the same signature as that of space-time (HDL clay **[41]**)
4) We will then compare on a large scale the behaviors of these materials on Earth with those of space-time in order to confirm or refute the foliated structure of space-time.

## 3. State of the art - theoretical framework and experimental results:

### 3.1. X-ray diffraction and lamellar structure

The observation of a crystal with X-rays consists of bombarding it with X-rays which, depending on the paths taken by the rays, will either collide with the atoms of the crystal's structure or pass through it without hindrance, drawing on the projection screen the very geometric structure of the crystal **[38]**, **[39]** and **[42]** to **[47]**.

X-ray diffraction is based on Bragg's law (see Figure 1), where $d_{hkl}$ is the spacing between atomic planes. The diffracted intensity depends on the crystal structure factor, the Lorentz-polarization factor, and the interlayer interference.

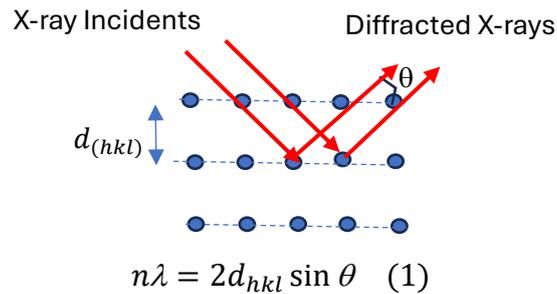

$$n\lambda = 2d_{hkl} \sin \theta \quad (1)$$

**Figure 1: Principle of Bragg's Law –**

It is then possible to draw X-ray diffractograms as a function of the angle of inclination 2θ of the rays and the intensity I of the spots on the screen **[41] [48]**. The number of peaks, their regular or irregular spacing, provides information on the leaf structure of the middle or not. The width of the peaks indicates the smallness of the structures. In crystals with a lamellar structure, multiple, regularly spaced 00ℓ peaks are observed (see figure 3), the analysis of which gives access to the thickness of the layers and their organization.

A typical example of a lamellar clay structure is given in Figure 2:

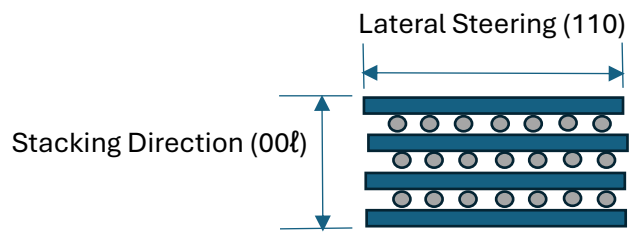

**Figure 2: Typical structure of HDL sedimentary clay with crystalline marking [41] [48] -**

An example of a lamellar structure diffractogram (HDL Clay) is given in Figure 3 below from the presentation **[41]** based on publication **[48]**.

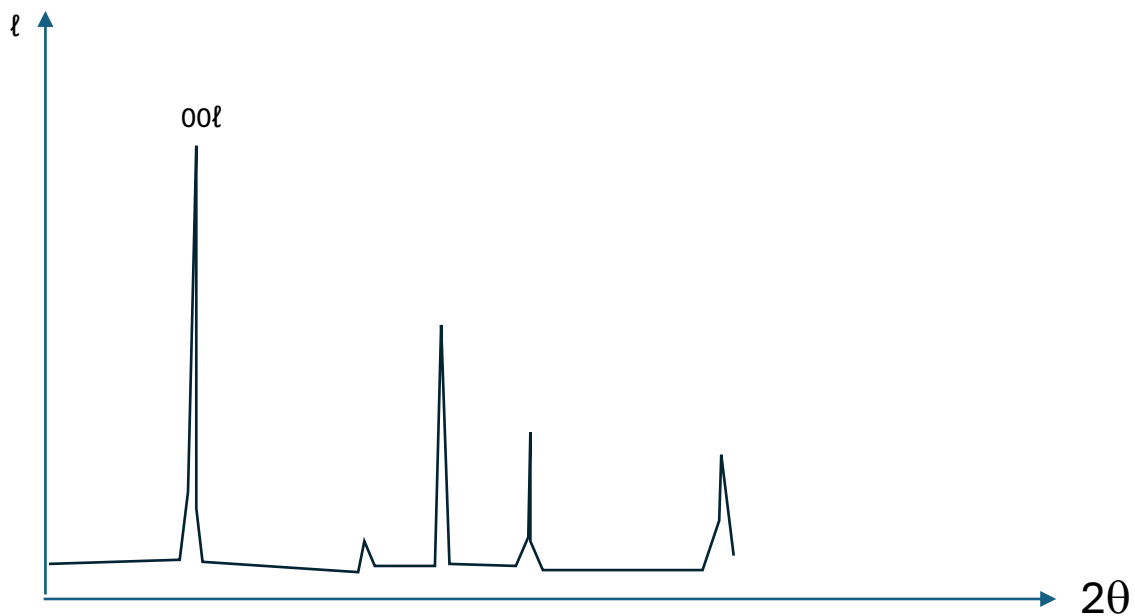

**Figure 3: Example of an X-ray diffractogram of HDL lamellar clay [41] [48]–**

The HDL clay diffractogram above shows in the 00ℓ direction a series of peaks that correspond to low angles to the harmonics associated with the direction of the stack (on the left of the diagram) and on the right for the large values of the angle, we see what is happening inside the sheets. The diffractograms above in Figure 3 are therefore typical of leaf structures.

We can also see in the figure 1 of **[41]** on the right the luminous intensities (equivalent to what we see in the case of the cosmic microwave background in temperature variation **[34]** to **[37]** but we will come back to this in the next chapter. The interpretation of a diffractogram is done in the following way:

The different peaks of a diffractogram are generally associated with the crystal periodicity of a material **[50]**. X-ray diffraction reveals the spatial structure of a crystal, indicating the distance between crystalline planes.

In the case of lamellar structures, the following can be noted:

      a) Multiple Ridges at Fixed Angles (2θ)

A lamellar structure can exhibit multiple peaks at fixed diffraction angles due to diffraction from parallel crystalline planes (lamellae). This can lead to multiple diffraction peaks at characteristic positions.

      b) The widening of the vertices in the direction perpendicular to the plane

If the lamellae are extended in the specific direction, this can lead to a widening of the diffraction peaks in the direction perpendicular to the crystalline plane.

      c) Anisotropic behaviors – peak-like anisotropy

The existence of preferentially aligned crystalline planes can lead to anisotropic behavior in the form of diffraction peaks.

      d) Peak intensity modulations

The presence of lamellae can cause periodic modulations in the intensity of diffraction peaks.

e) Additional Thoughts

Additional reflections (apart from the main reflections) may occur due to diffractions of the interlamellar planes. The analysis of peaks is generally done as follows:

If we apply all the above to the case of lamellar clays, we can highlight the specificities of the associated diffractogram.

### a) Concerning the Spokes 00ℓ :

In the presence of parallel sheets in a multilayer material, 00ℓ lines are observed in the diffractogram. The position of these lines will depend on the interplanar distances associated with the crystalline planes formed by the different layers, which in turn are made up of sets of atoms distributed in planes.

### b) Concerning the series of peaks:

Regularly spaced peaks, each corresponding to a specific order of diffraction n and a particular interplanar distance, will appear. These peaks are usually associated with constructive diffraction in parallel crystalline planes.

### c) Regarding Relative Currents on the control screen:

The relative intensity of the peaks will depend on the density of the atoms in the different shells and the distribution of electrons. Variations in intensity may indicate differences in the composition or arrangement of the layers.

### d) Concerning the shape of the peaks:

The shape of the peaks can provide indications of the quality of the crystals, the size of the crystals as well as the possible presence of defects or imperfections in the structure.

### e) Concerning the width of the peaks:

The width of the peaks is influenced by the size of the crystals, especially the imperfections of the crystal. So, if we assume that the power spectra of the cosmic microwave background are some kind of inverted X-ray diffractogram (see Figure 4), and if we analyze the peaks of the CMB as if it were an inverted X-ray diffractogram, we should also see more or less regularly spaced peaks.

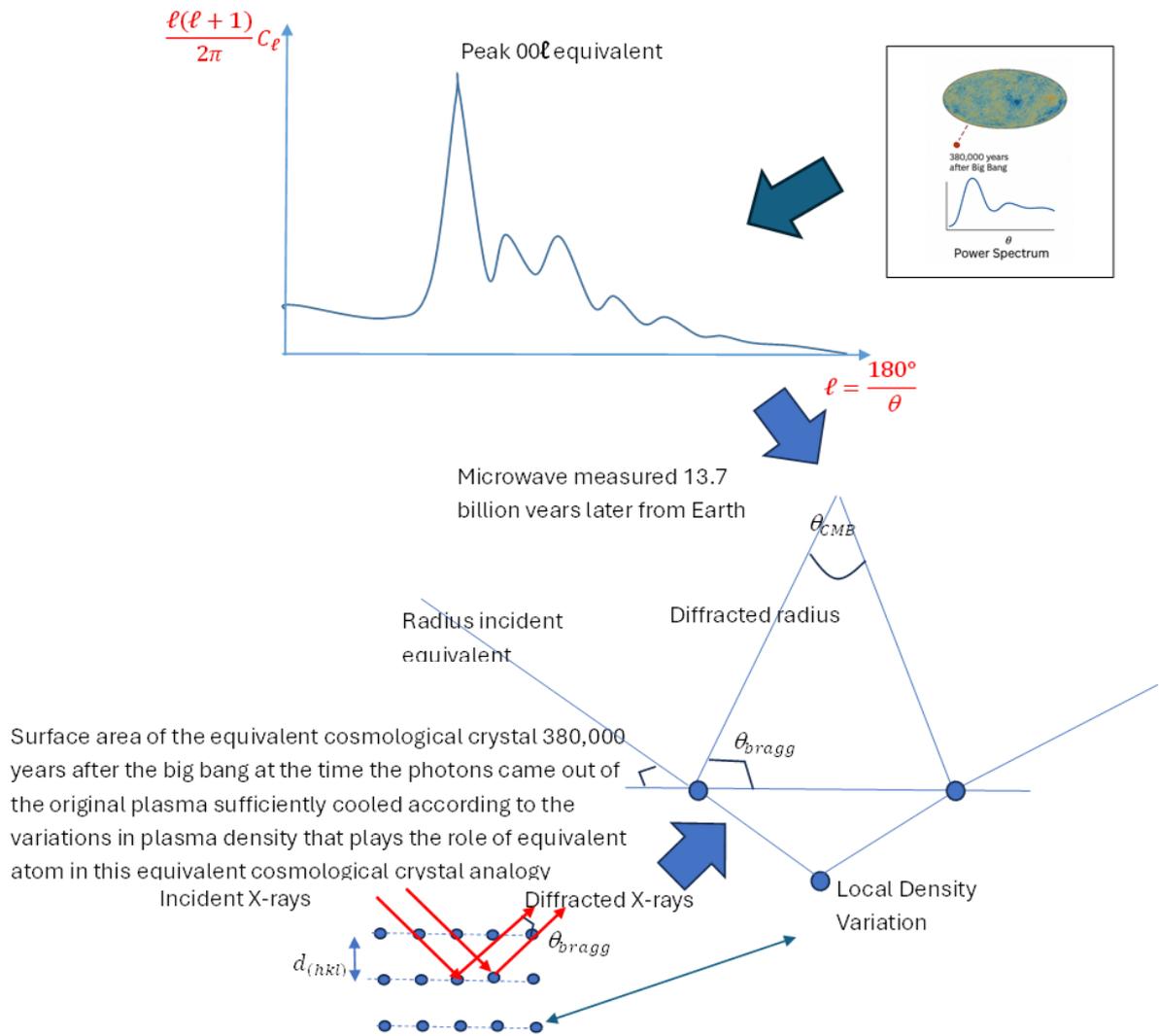

**Figure 4: Visualization of the inverted diffractogram of space-time 380,000 years after the big bang -**

## 3.2. CMB angular power spectrum

### 3.2.1 General and principle of power spectra and results obtained

The CMB gave rise to two types of power spectra **[34]** to **[37]**, **[49]**. The first relates temperature variations to the angular multipole $\ell$. The second connects the B or E type polarizations.

The power spectrum is defined by the average variance of the coefficients deriving from the spherical harmonic decomposition of the temperature field ΔT. The multipole $\ell$ is inversely proportional to the angle of observation. It is important to note that the spectrum obtained for temperature corresponds exactly to the blackbody spectrum predicted in cosmology.

The peaks observed correspond to the fundamental and harmonic modes of the acoustic oscillations of the primordial photon-baryon plasma **[34]** to **[37]** and **[51]**.

The study of this cosmic microwave background by the Wmap **[52]** and Planck **[34]** to **[37]** satellites has given rise to three main results.

- The first concerns the infinitesimal temperature variations as a function of the aperture of the angle of observation. This is the temperature power spectrum (see Figure 3). The curve obtained corresponds with great precision to the theoretical curve, thus validating the big bang model,
- The second concerns the polarizations of photons as a function of deformations **[53] [54]** and therefore of the structure and nature of the original plasma 380000 years after the big bang according to two modes E and B (see figure 4). Polarization spectra for each of these modes were also established. They also correspond with great precision to the theoretical spectra from the big bang theory.
- Third, the universe has almost zero curvature (see **[36]** and **[55]** to **[58]**).

### 3.2.2 Temperature Power Spectrum Results

The curve of the power spectrum obtained for the universe 380000 years after the big bang is given in Figure 5 below. It follows the theoretical curve of the perfect blackbody perfectly.

From a mathematical point of view, spherical harmonics are obtained from the following well known mathematical factors:

$$C_\ell = \frac{1}{2\ell+1} \sum_{m=-\ell}^{m=+\ell} |a_{\ell m}|^2 = \langle a^*_{\ell m} a_{\ell m} \rangle = \langle |a_{\ell m}|^2 \rangle_m \quad (2)$$

The expression $a_{\ell m}$ for the integration of the solid angle over the whole sphere (4π) is written:

$$a_{\ell m} = \int_{4\pi} \frac{\Delta T_{(\theta,\varphi)}}{T} Y^*_{\ell m(\theta,\varphi)} d\Omega \quad (3)$$

And $Y^*_{\ell m(\theta,\varphi)}$ is deduced from the infinitesimal temperature variations T of the cosmic microwave background:

$$\frac{T_{(\theta,\varphi)} - T}{T} = \frac{\Delta T_{(\theta,\varphi)}}{T} = \sum_{\ell=0}^{m} \sum_{m=-\ell}^{\ell} a_{\ell m} Y_{\ell m(\theta,\varphi)} = \sum_{\ell=0}^{m} \sum_{m=-\ell}^{\ell} a_{\ell m} \sqrt{\frac{(2\ell+1)(\ell-m)!}{4\pi(\ell+m)!}} P_{\ell}^{m}(\cos\theta) e^{\pm im\varphi} \quad (4)$$

Figure 5 below shows the typical shape of the power spectrum in temperature.

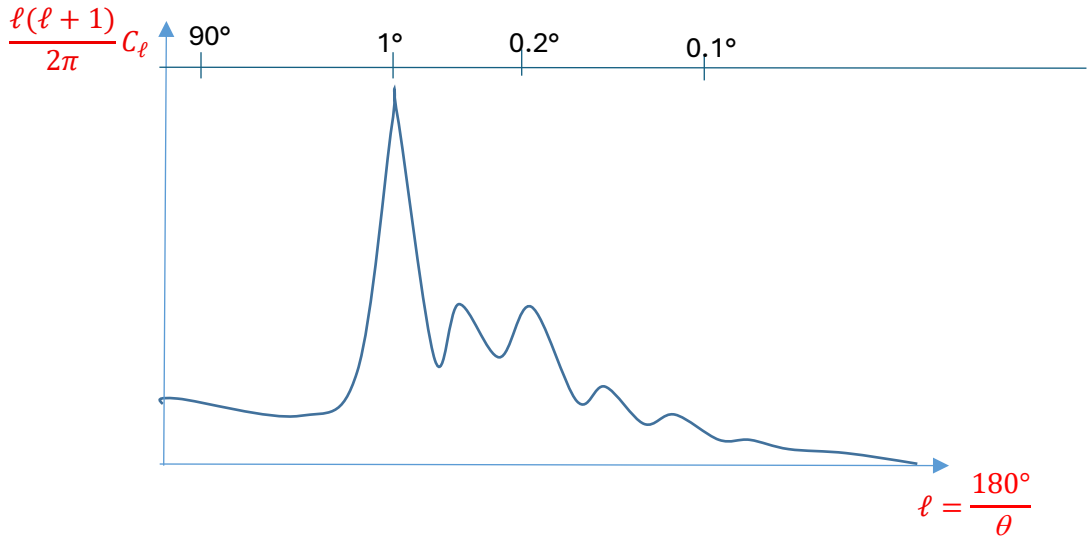

**Figure 5: Cosmic microwave background power spectrum (wide angles on the left and small angles on the right) [54] [59] $(\mu K)^2 -$**

The interpretation of the different peaks by cosmology is as follows **[59]**:

The angular scale of the first peak is usually related to the curvature of the universe. The second peak, related to the ratio of odd to even peaks, is used to determine the reduced density of baryons. The third peak is related to the calculation of the density of dark matter (a dimension much smaller than the red and blue spots of the CMB) which sculpts the formation of galaxies (constant speed of the speeds of stars near the central black hole and on the periphery in contradiction with Newton's law) or the formation of the cosmic web (filamentary organization of galaxies that follow channels, curvature of dark matter). A more accurate reading of the peaks of the cosmic microwave background can be made in the following way, according to **[59]**.

For the first peak corresponding to acoustic resonance (see Figure 6):

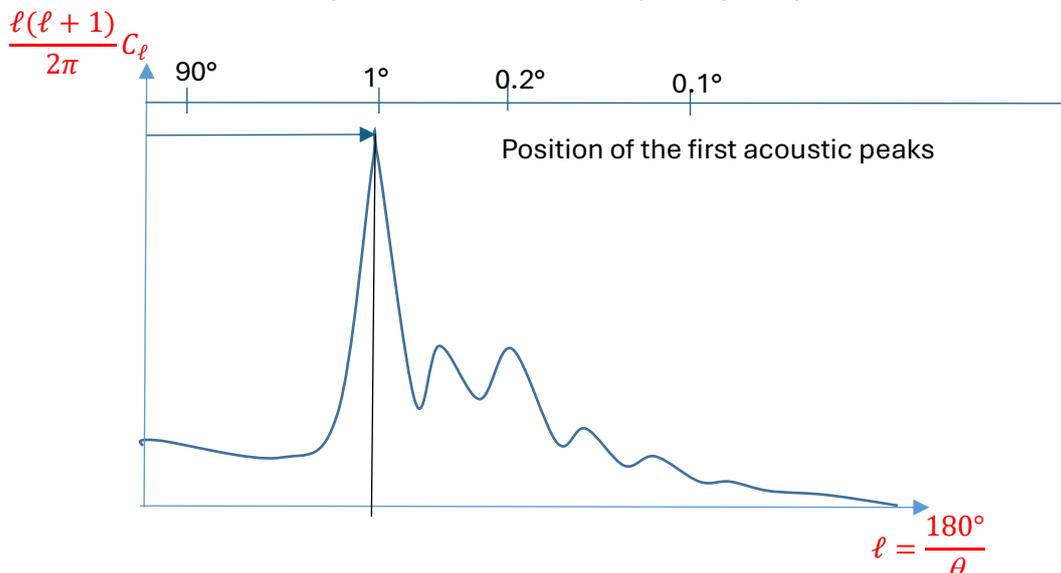

**Figure 6: Interpretation of the first peak of the power spectrum of the source [59] -**

Following **[59]** "The position of the first peak of the FDC curve (about 1°) corresponds to the angular distance below which we see today the fundamental wavelength of the oscillations - the distance traveled by the "sound" in overdensities in 380000 years!!"

This position also depends on the curvature of the universe and its expansion rate (i.e., dark energy).

The oscillations that generated overdensities (wavelengths 2, 3, 4… times greater than the fundamental wavelength) – the 2nd, 3rd, 4th peaks of the cosmic microwave background should correspond to them. We will come back to this, but the regular spacing of these harmonics reflects a regulation of vibrations and therefore of the structure of the primordial universe; 00l repetitive peaks at regular intervals of leaflet crystals in X-ray diffractograms.

The successive peaks reflect successive compressions and relaxations in the original plasma (see Figure 7).

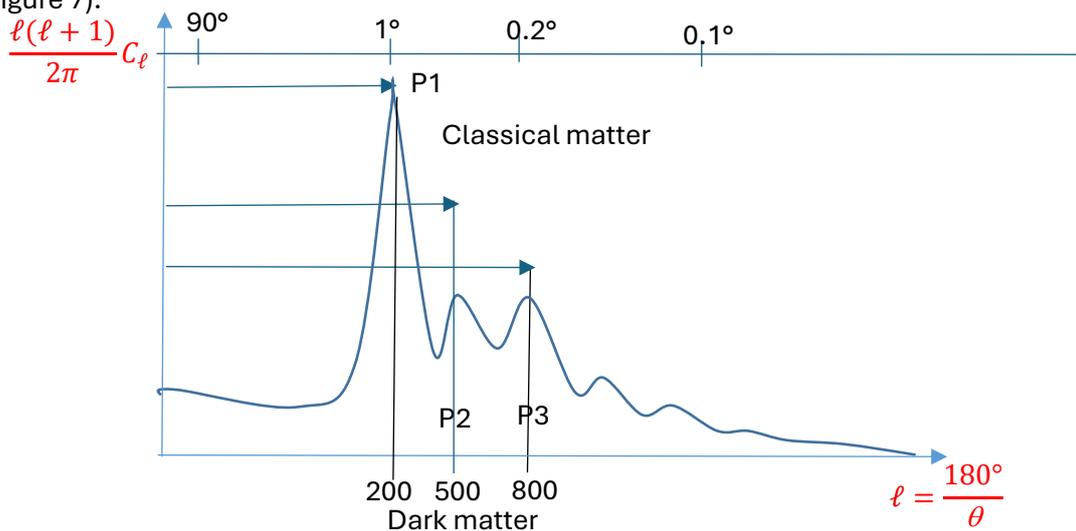

**Figure 7: Interpretation of harmonic peaks in the power spectrum of the source [59]-**

The measurement of the P1, P2, P3 ratios gives access to the classical matter/dark matter ratio.

Next **[59]** "The higher the local density of classical matter, the more intense the "compression"; Hence the relative increase in the intensity of odd peaks and the relative decrease in the intensity of even peaks.

A high local density of dark matter is manifested by an increase in the intensity of odd peaks.

From the measurement of the intensities of the first three peaks, we finally derive information on:

- The curvature of the universe

- the amount of dark energy

- the amount of conventional material

- the amount of dark matter"

From a light intensity point of view, the CMB gives rise, as for the X-ray diffractogram, to shades of color as a function of temperature variations associated with variations in material density, as shown in Figure 8 below.

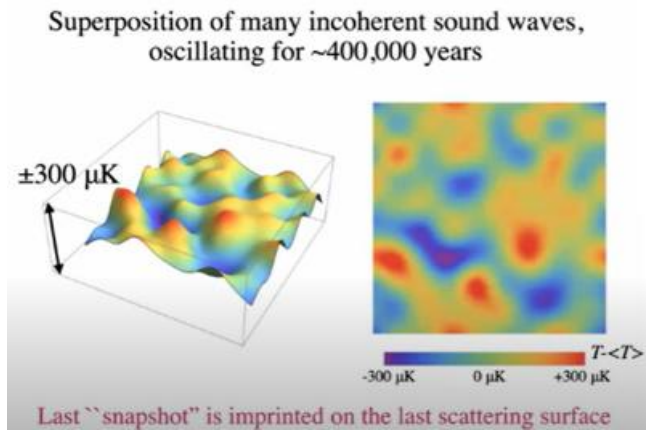

**Figure 8: Zoom on a temperature variation in the CMB [60] -**

### 3.2.3 Results of the polarization power spectrum.

During the emission of photons, plasma and dark matter exhibited rotational dynamics. Type B and Type E polarizations.

This phenomenon is well summarized in "February 22, 2022| FOR SCIENCE N° 533" I quote:

In the primordial plasma, ordinary matter interacts strongly with photons. The antagonistic forces (the gravitational force of matter, which is attractive, and the radiative pressure of photons, which is repulsive) caused matter to oscillate, producing areas of high density and others of low density, such as "acoustic waves". With these ingredients alone, it is not possible to reproduce these fluctuations satisfactorily. In addition, there is an element such as dark matter, which has accentuated the formation of overdensities. Since it is subject only to the gravitational force, it has accumulated in some regions. The latter, creating even more intense gravitational wells, attracted more ordinary matter.

The typical polarized power spectra are given in Figure 9 below.

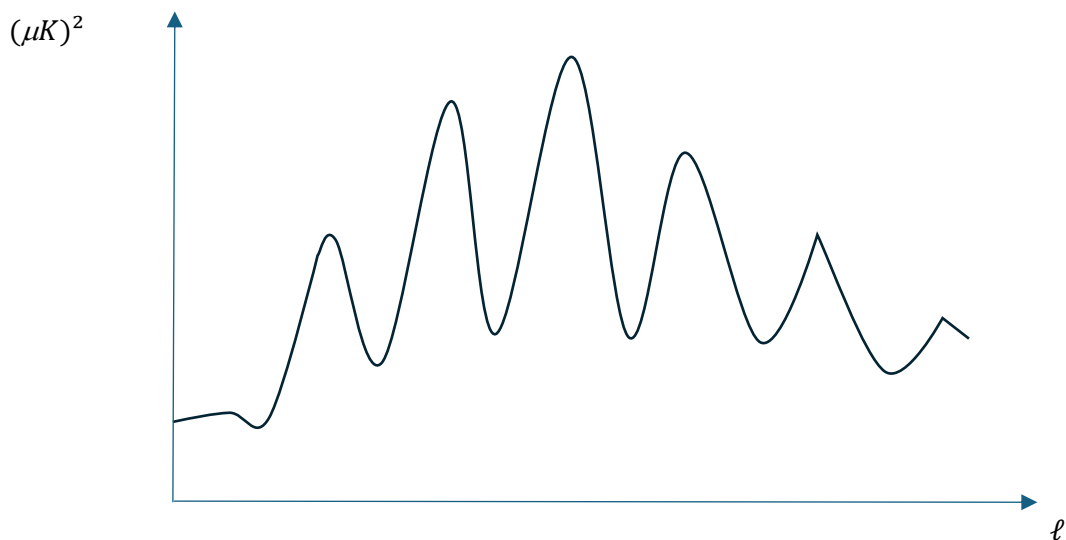

**Figure 9: E-mode polarized power spectrum [49]-**

Remark

The shape of the polarizations can be related to the shape of the structures through which they pass according to the principle below (see Figure 10). The wave that passes through is the one that is in a direction not blocked by the network. In our case, it is the structure of space (dark matter?) acting as a grid (clusters of baryons, neutrinos, dark matter) that allows the waves to pass more or less in E or B mode depending on their organization in the original plasma. The two modes are shown in Figure 10 below

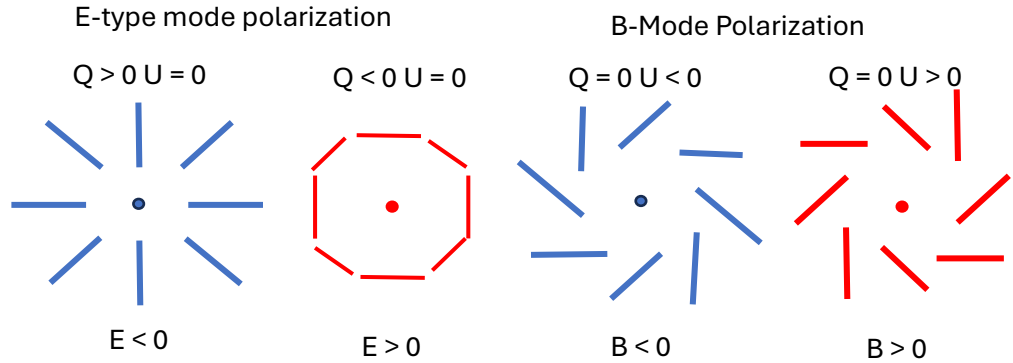

**Figure 10: View of the different polarizations of the E and B modes of the cosmic background source [54]** —

## 4. Expected results and discussion

### 4.1 General

We have seen that in the case of the Cosmic Microwave Background (CMB), the equivalent X-ray diffractogram is in some way inverted (Figure 4). Light rays are created in the crystal by the energy of the big bang (in X-rays they are produced outside the crystal). They come out of it influenced by the very structure of the universe 380000 years after the big bang (in this case the space-time filled with plasma distributed according to its structure and internal dynamics, the distribution of which depends on the structure of space-time itself, the density of dark matter and baryonic matter, etc.). Depending on these structures through which the photons pass, this creates temperature fluctuations (the counterpart of intensity I fluctuations in X-rays) which gives us an image of the structure of plasmas and their polarizations (how they rotate). Depending on the angle of observation on the x-axis of the power spectra (T or energy, θ), we see different structures via the opening distribution of the different peaks as in the case of Bragg's law and the associated X diffractograms. From these peaks, as a function of the angle of observation of the light rays, we deduce the internal structure of space-time at that time. We can thus see the analogy between the two graphs as shown in tables 1 in temperature and 2 in polarizations below. The similarities of the analogy are clearly visible.

In view of the previous state of the art, if we admit that the power spectra are the equivalent of the X-ray diffractogram of the space-time crystal, we expect to find:

- A marked periodicity of the peaks (analogous to the $00\ell$ peaks in a lamellar crystal), to be in agreement with the foliated space described in the case of gravitational wave analysis **[24]**, (the space-time today resulting from the expansion of space-time at the time of the emission of the CMB),
- A large first line, reflecting the initial dispersion of the sizes of small "grains" relatively speaking in relation to the size of the current universe, (in agreement with the hypotheses of T Tenev and M.F Horstemeyer **[12] [13]**.
- A gradual decrease in the relative intensity of harmonics, testifying to the dynamic evolution of the plasma.

We'll also discuss the subtle effects:

- Possible torsion of space-time detected by a shift of the BB modes with respect to the EE modes,
- Influence of dark matter density on peak regularity,
- Constraints on the curvature of the universe via the position of the first peak.

**4.2 Comparative analysis of the temperature power spectrum with typical X-ray diffractograms of lamellar clay**

Table 1 below allows us to visualize the similarity in the peaks repetitions between the lamellar clay structures **[41] [48]** and between the temperature power spectra of the CMB **[34]** to **[37] [49] [60]**. It is also observed that the luminous spots of varying intensity of the lamellar clays are found in the same way in the temperature variations in the cosmic microwave background.

| The case of lamellar HDL clays under X-ray diffractogram | Case of the power spectrum of the cosmic background temperature |
|---|---|
| Harmonics associated with the direction of stacking of atomic planes figure 2 of **[41]**. 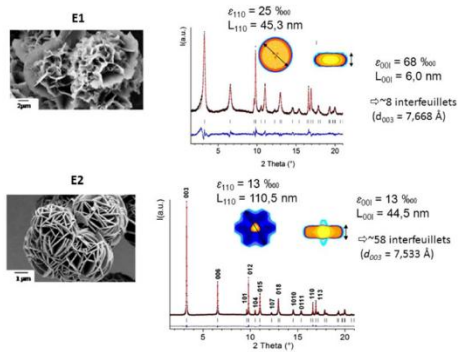 Light Spot Intensity figure 1 of **[41]** 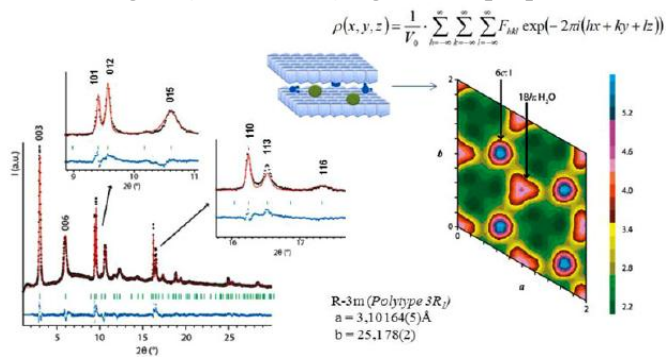 | Harmonics associated with density variations in the original plasma **[60]**. 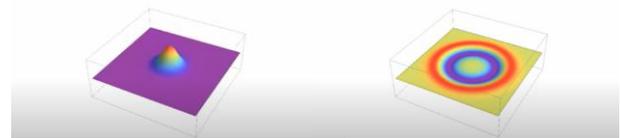 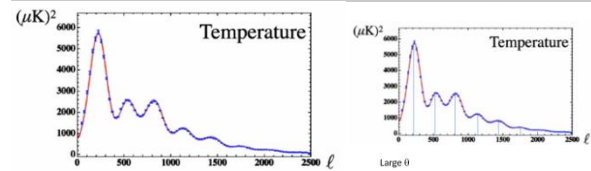 Intensity of the thermal variation visualized with a light point **[60]** 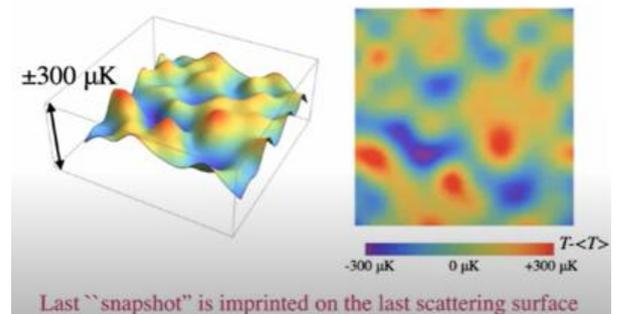 Mechanism in the original plasma (the variation in velocity in the plasma generates polarizations and temperature variations **[60]** |

**[41]** Credit:EDP sciences : "Étude du mécanisme d'échange et de la structure des matériaux hydroxydes doubles lamellaires (HDL) par diffraction et diffusion des rayons X – 2013 - UVX2012 - 11e Colloque sur les Sources Cohérentes et Incohérentes UV, VUV et X ; Applications et Développements Récents - https://doi.org/10.1051/uvx/201301016 - C. Taviot-Guého and all. »

**Table 1: Analogy between the measurements of the thermal vibrations of atoms in the case of the X-ray diffractogram and the temperature variations at the time of the release of photons in the source cosmic diffuse field [41], [60] —**

The analysis of the power spectrum of the cosmic microwave background (Table 1) from this angle reminds us of the X-ray diffractograms of lamellae clays **[41] [48]** given the regularity of the peaks. Indeed, the maximums of the peaks of the CMB are , $\ell_1 \approx 220, \ell_2 \approx 540, \ell_3 \approx 800, \ell_4 \approx 100$ or the successive deviations between each consecutive peak (So, if the power spectrum of the CMB is read as an inverted X diffractogram, then we find a certain regular foliation of space-time at the time of the CMB mission. $\ell_2 - \ell_1 = 320, \ell_3 - \ell_2 = 260, \ell_4 - \ell_3 = 300$ .

Moreover, the peaks being narrow, this implies a fine granular structure, as in the case of clays, all things being equal, of course, i.e. the size of the plasma clusters at the time of the big bang, which are very small compared to the size of the universe today. It is interesting to see that here again we find the structure that emerged during the analysis of gravitational waves with and without geometric torsion, confirming the validity of the models of space with a leaf structure developed in ADM **[33]**, Tenev and Horstemeyer **[12] [13]** and Izabel et al **[24]**.

**4.3 Proposal for a cosmological Bragg law**

**4.3.1 Classical Bragg's Law (crystallography)**

We saw in the previous paragraph, that if we analyze the cosmic microwave background as an inverted X-ray diffractogram (figure 4) (we do not bombard a structure from the outside with X-rays, they are light rays and therefore energetic photons, which pass through a structure from the inside to the outside), then we find in the distribution of the regular different successive peaks a certain foliated structure of space. This analogy seems qualitatively well founded, since in the X-ray diffractogram we have a diffraction angle of 2θ on the x-axis, and in the case of the CMB a relative angle $\ell = \frac{180}{\theta}$ on the one hand, and on the y-axis, a luminous intensity of the X-rays in the X-ray diffractogram and a temperature variation attributable to an energy in the cosmic microwave background on the other hand. In addition, we have a crystal structure as a physical object in the case of X-ray analysis on the one hand, and acoustic oscillations necessarily attached to an underlying structure in vibration (pressure, local decompression of baryon plasmas as a function of dark matter concentrations, see figures 8 and 13) in the CMB on the other hand. However, for this quantitative analogy to take on its full meaning, it would be necessary to qualitatively find in the case of the CMB a certain form of cosmological Bragg's law that would somehow validate this analogy. This is what we will study in this paragraph.

In a crystal, electromagnetic waves (X-rays) are diffracted by atomic planes spaced at a distance d. The condition for obtaining constructive interference (a peak of intensity on the diffractogram) is given by Bragg's law **[42]** to **[47]**:

$$n\lambda = 2d\sin\theta \quad (5)$$

where:

- n is the order of diffraction,
- λ is the incident wavelength of X-rays,
- d is the atomic interplane distance of the structure of the crystal studied.
- θ is the angle of diffraction of the rays.

At the atomic scale, the X-ray wavelength is well suited to see the atomic structures of the crystals studied.

### 4.3.2 Cosmological correspondence of Bragg elements

In the framework of the Cosmic Microwave Background (CMB), based on the state of the art established in paragraph 3 and on the basis of our preliminary analysis in 4.3.1, we propose a direct analogy between X-ray diffraction and cosmology.

The principle of analogy is given in Figure 10 below in the case of the cosmic microwave background. Each time the photon passes through a "sheet" of space (a shell), it corresponds to a peak in the CMB power spectrum, exactly as in X-ray diffraction of a lamellar clay **[41]**, where each plane produces a peak at a characteristic angle. This visually confirms the foliation of space at the acoustic scale, with concentrated "shells" and regular angular peaks in the spectrum. Diagram 10 below shows two elements side by side:

1. Left: trajectory of a photon (in orange) starting from the observer, passing through three spherical shells (in black) each spaced by $r_s \approx 0.145$ Gpc **[36]** ($r_* = r_s(z_*)$) with $r_s$ the sound overpressure zone). The red dots mark the intersections of the photon with each shell.
2. Right: Simplified angular power spectrum. The vertical dotted lines indicate the angles $\theta_n$ corresponding to the intersections:
   - First peak at $\theta_1 \approx \frac{180}{220} = °0.60$,
   - Second peak at $\theta_2 \approx \frac{180}{540} = °0.33$,
   - Third peak at $\theta_3 \approx \frac{180}{800} = °0.22$.

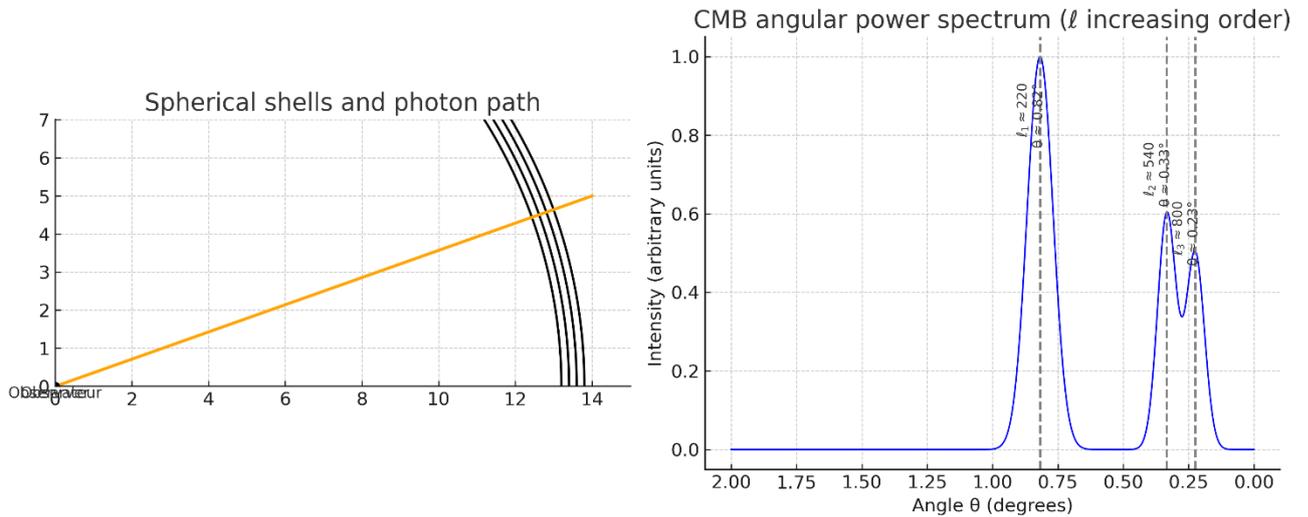

**Figure 11: Principle of the equivalent cosmological crystal analogy and of an "equivalent inversed cosmic diffractogram X"-**

This analogy is therefore detailed at the parameter level in Table 2 below. The definition of the different characteristics is given in Figure 1 for crystallography and Figure 11 for cosmology.

| Crystallography | Cosmology |
|---|---|
| Atomic Plan | Acoustic shell (spherical space sheet) |
| Atom / Mesh | Size sound boost zone $r_s$ |
| Incident X-ray | Photon du CMB |
| Detector | Observer (today) |
| Bragg angle θ | Angle θ at which a spot of the CMB is observed |
| Detector-to-sample distance | Comoving angular distance between observer and acoustic shell $D_A$ |

**Table 2: Correspondence by analogy between crystallography and cosmology (cosmic microwave background) –**

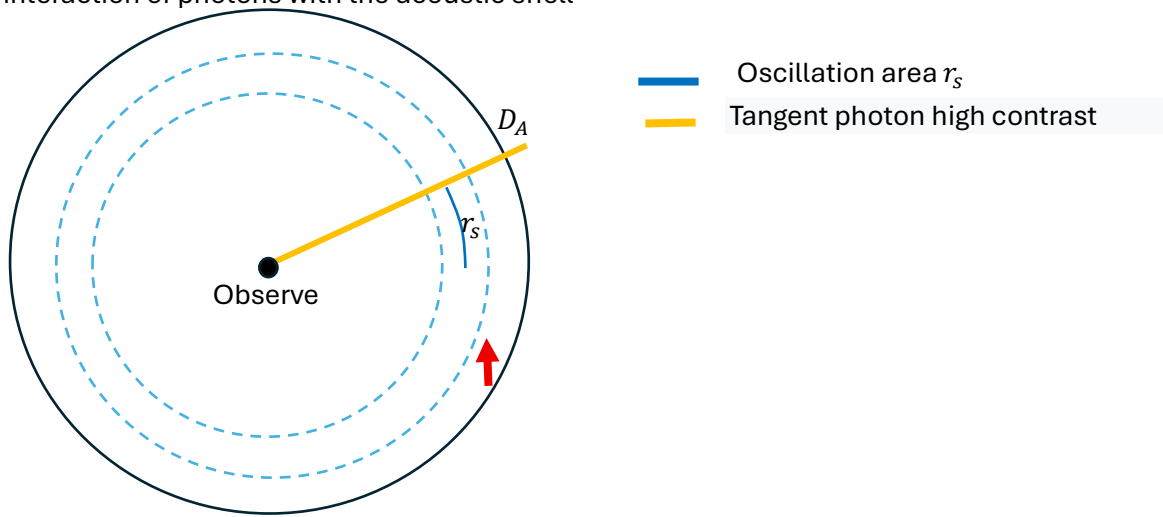

**Figure 12: Representation on the left of the set of parameters of the analogy in spherical shells -**

In Figure 12, the interaction of the photons with the acoustic shells of the CMB has been shown on the left. The central circle is the observer. The spherical shells (in black and dotted) represent the acoustic space sheets separated by $r_s$. The blue arc marks an area of acoustic oscillation visible at an angle θ. That is to say, the dimension of one of the red or blue spots of acoustic oscillation of the plasma related to the density of baryon and locally dark matter that can be seen on the Planck map of the CMB (see figure 13)

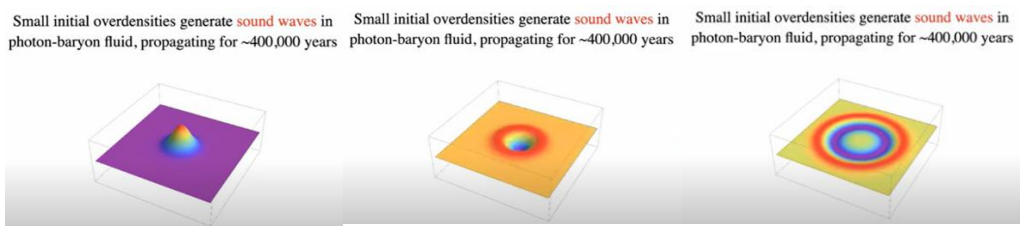

**Figure 13: Example of dimensional acoustic oscillation $r_s$ at the base of the color spots of the cosmic microwave background [60] -**

Figure 12 shows two types of photon trajectories:

- o Red (radial): crosses perpendicularly a layer that gives low contrast (little peak).
- o Orange (tangential): cuts the layer for a long time, which gives strong contrast (intense peak).

Thus, the orientation of the photon ray with respect to the acoustic shells determines the intensity of the peak in the spectrum of the CMB. Its angular width (more tangent gives a sharper image) and therefore we draw information about the geometry of the primordial universe.

### 4.3.3 Bragg's Law Transformed into Cosmology

To go beyond the purely analogical framework, we propose here an effective mechanical interpretation: the acoustic peaks of the CMB are seen as the eigenmodes of a cohesive stratified medium, whose structure is modeled by a superposition of spherical shells (acoustic shells). This structure is compatible with ADM, Einstein–Cartan and continuous media theories.

On a large scale, the universe can therefore be seen as a sphere of CMB radiation, where photons have passed through a set of laminated spherical shells (Figure 11) (the equivalent of the atomic planes of a crystal) with dimensional acoustic vibrating zones $r_s$ (which are the counterpart of atoms in a crystal tested by X-ray). Depending on the angle at which the photons interact with its equivalent atoms, peaks appear on the power spectrum. If our reasoning and analogy is correct, we should be able to find the counterpart of a cosmological Bragg law. That is the purpose of this paragraph. By taking Bragg's law and stating:

- $\theta_n = \frac{\lambda_n}{D_A}$ the angle that the wave covers today,
- $\ell_n = \frac{2\pi}{\theta_n}$ spherical harmonic of degree $\ell_n$,
- $\lambda_n = 2\pi D_A/\ell_n$ (acoustic wavelength of order n), $2\pi$ coming from the frequency/harmonic wavelength correspondence in sphere
- $d = r_s$ if the acoustic horizon is the same size as the interhull spacing
- $\theta = \theta_n =$ (angle at which a multipole peak is observed), the apparent angle of the nth peak$\ell_n$

We obtain Bragg–CMB's law from the well-known Bragg's law:

$$n\lambda_n = 2d \sin\theta_n \quad (6)$$

Either:

$$n\frac{2\pi D_A}{\ell_n} = 2r_s \sin(\theta_n) \quad (7)$$

In X-ray diffraction, the most intense peak occurs when the crystalline planes are orthogonal to the direction of the diffracted radius, i.e. for θ = 90° (sin θ = 1). In addition, in the CMB, acoustic oscillations are described by k-wave vector modes. The temperature anisotropies that we measure come mainly from modes whose wave vector is perpendicular to the line of sight (µ = cos angle = 0):

- These modes do not move photons either to or from the observer,
- They produce a variation in maximum density visible as a spot on the sky.

This condition µ=0 corresponds exactly to a 90° diffraction in our crystal-equivalent.

Indeed, the peaks in the power spectrum come from $C_\ell$ the acoustic modes that were in the maximum compression phase at the time of decoupling. These modes have a wave vector k, such that $\mu = \cos(k, n) = 0$ i.e. orthogonal to the line of sight. This geometry is equivalent, in our crystallographic analogy, to a maximum Bragg angle diffraction (θ=90°), i.e. sinθ=1.

Therefore:

$$sin\theta_n = 1 \quad (8)$$

We obtain:

$$n\frac{2\pi D_A}{\ell_n} = 2r_s \quad (9)$$

We therefore obtain Bragg's law generalized to the cosmological crystal:

$$\ell_n = n\frac{\pi D_A}{r_s} \quad (10)$$

We will test it digitally in the next paragraph.

### 4.3.4 Numerical Application (with Data Justification)

To do this, we use experimental data (Planck 2018) recalled in Figure 14.

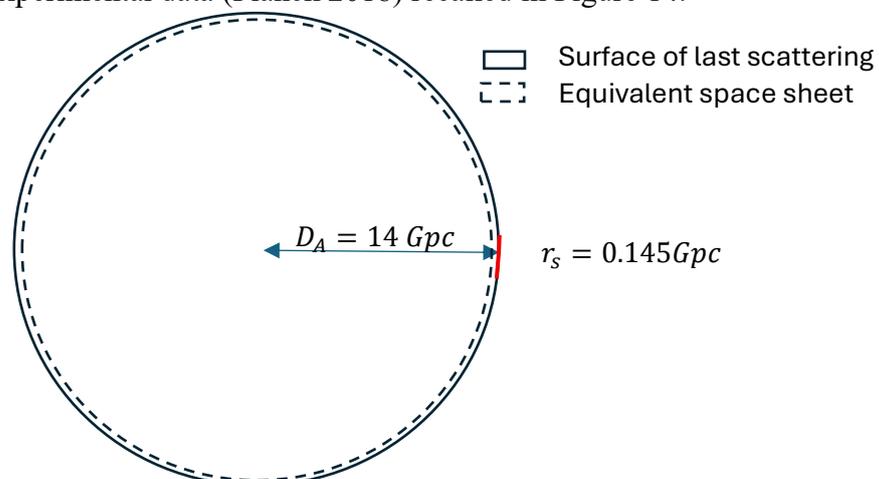

**Figure 14: Different data of the cosmic microwave background from the Planck experiment in 2018-**

- The co-moving acoustic horizon: $r_s$ =144.4 Mpc according to **[36] [62]**.
- We know that the distance that separates us today from the surface of the last scattering is about 43 billion light-years. That is $D_A$ =13,897 GPc **[62]**

Applying the formula $\ell_n = n \frac{\pi D_A}{r_s}$ with the above data yields the results in Table 3 below.

| Order n | $Formula$ | Calculated value | Observed peak (Planck) | Corrected peak |
|---|---|---|---|---|
| 1 | $\frac{\pi D_A}{r_s}$ | ≈302 | $\ell_1$≈220 | $\ell_1$≈304±4 (*) |
| 2 | $2\frac{\pi D_A}{r_s}$ | ≈604 | $\ell_2$≈540 | - |
| 3 | $3\frac{\pi D_A}{r_s}$ | ≈907 | $\ell_3$≈800 | - |
| 4 | $4\frac{\pi D_A}{r_s}$ | ≈1209 | $\ell_4$≈1100 | - |
| (*) According to publication **[51]** | | | | |

**Table 3: positions of the first 4 peaks calculated from the cosmological Bragg's law applied to the equivalent cosmic crystal compared to Planck's experimental results in 2018**

Although the first acoustic peak of the CMB is often quoted at $\ell \approx 220$, this value corresponds to the local maximum of the temperature power spectrum. However, due to the asymmetric shape of the peak—caused by Silk damping, projection effects, and plasma composition—the true spectral centroid lies closer to $\ell \approx 304 \pm 4$, as shown in **[51]**. This corrected value aligns remarkably well with our prediction from the cosmological Bragg law ($\ell_1 \approx 302$), reinforcing the validity of the analogy."

We can therefore see that unlike the Standard Model which adjusts peaks via Boltzmann solvers, our approach directly predicts the positions of the peaks from a simple geometric law. This prediction, independent of the adjusted cosmological parameters, constitutes a form of complementary explanatory power.

**CMB Acoustic Peak Comparison**

This figure 15 compares the predicted positions of the first four acoustic peaks in the Cosmic Microwave Background (CMB) using the cosmological Bragg law with the observed values from the Planck satellite. The corrected barycentric value of the first peak ($\ell_1 \approx 304 \pm 4$) is also shown for comparison.

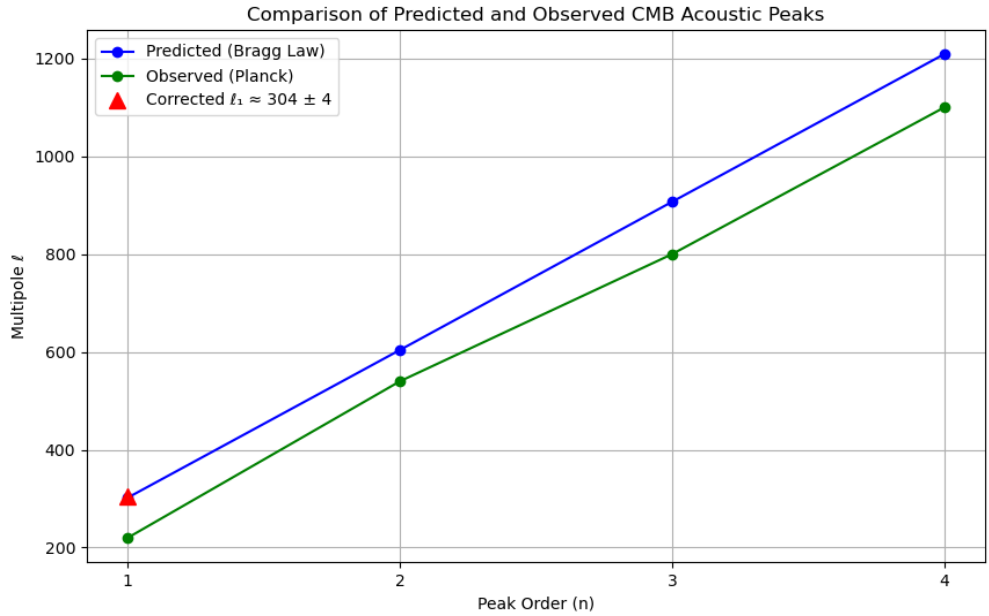

**Figure 15: Comparison of predicted and observed CMB acoustic peak positions.**

Legend:

- **Blue lines**: values predicted by the cosmological Bragg law (ℓ ≈ 302, 604, 907, 1209)
- **Green lines**: peaks observed by the Planck satellite (ℓ ≈ 220, 540, 800, 1100)
- **Red triangle**: corrected barycentric value of the first peak (ℓ₁ ≈ 304 ± 4)

The predicted peak positions from the cosmological Bragg law systematically overshoot the observed Planck values by approximately 9–11%.

- ℓ₁ corrected ≈ 304 → very near of ℓ₁ prédicted ≈ 302
- ℓ₂ observed = 540 vs ℓ₂ near= 604 → ratio ≈ 0.89
- ℓ₃ observed = 800 vs ℓ₃ prédicted = 907 → ratio ≈ 0.88
- ℓ₄ observed = 1100 vs ℓ₄ prédicted = 1209 → ratio ≈ 0.91

This discrepancy can be interpreted as a projection or curvature correction factor, possibly arising from residual baryonic loading, geometric projection effects, or torsional distortions in the primordial spacetime. Applying a global correction factor of ≈ 0.89 to the Bragg-predicted values aligns them remarkably well with the observed spectrum, reinforcing the structural validity of the analogy. This coefficient could be justified by:

### A) Residual Curvature Effects

Even though the universe is globally flat, a local effective curvature at the time of decoupling could have altered the angular projection of the acoustic peaks.

### B) Spherical Projection Effects

The cosmological Bragg law is derived within a simplified geometry. A geometric correction related to the projection onto the celestial sphere could account for this factor.

### C) Baryonic Density Effects

As mentioned in the text, baryonic density and diffusion effects (Silk damping) shift the peaks to the left. This factor could be interpreted as an effective baryonic correction.

### D) Analogies with Real Crystals

In real crystals, diffraction peaks can be shifted due to defects, internal stresses, or density gradients. A parallel can be drawn with residual torsion or a dark matter density gradient in the primordial universe.

### E) Geometric Torsion effect in the Crystal

In the Einstein–Cartan framework, geometric torsion modifies the propagation of acoustic and electromagnetic waves by introducing additional degrees of freedom and phase shifts. These effects can lead to an effective angular compression of the observed CMB peaks. A global correction factor of $\chi \approx 0.9$ may thus reflect the cumulative influence of torsional distortions on the angular diameter distance and the harmonic structure of primordial spacetime.

The systematic offset between the predicted and observed positions of the CMB acoustic peaks—quantified by a global correction factor of approximately 0.9—may be interpreted as a geometric signature of spacetime torsion. In the Einstein–Cartan framework, torsion modifies the propagation of acoustic and electromagnetic waves by introducing additional degrees of freedom and phase shifts. These effects can lead to an effective reduction in the angular diameter distance, thereby compressing the observed angular scales. Recent theoretical work (Wu et al., 2024 ) **[71]** shows that torsion alters the Friedmann dynamics and can shift cosmological observables such as the Hubble constant and distance measures. While the study does not explicitly quantify the angular compression of the CMB peaks, it supports the idea that torsion can mimic curvature-like effects and reduce the effective angular scale—consistent with the 10% compression observed in our Bragg-based model.

**Conclusion**

The analogy between the cosmic microwave background and a periodic crystal is not only conceptually rich, but also quantitatively robust. Indeed :

1. Concerning the acoustic resonance conditions.

    The oscillations of the primordial plasma satisfy the condition $k\,r_s = n\pi$, which, once translated into multipoles via $k \approx l/D_A$, naturally leads to:

$$\ell_n = n\frac{\pi D_A}{r_s} \quad (11)$$

2. Regarding Bragg's Law–CMB

   By interpreting the acoustic shells as the planes of a "cosmic" crystal, and by applying an adapted version of Bragg's law with $sin\theta_n = 1$ (transverse modes), we find the same relationship $\ell_n = n\frac{\pi D_A}{r_s}$, confirming the coherence of the analogy.

   Regarding experimental validation

   With $D_A \simeq 13{,}900$ MPc [62] and $r_s \simeq 144$ Mpc [36] and [62], the law $\ell_n = n\frac{\pi D_A}{r_s}$ gives:

   $\ell_1 \approx 302$ et $\ell_2 \approx 604$, $\ell_3 \approx 907$, $\ell_4 \approx 1209$,

   in good agreement — with a correction factor related to baryon densities and curvature — with the structures observed in the Planck spectrum [34] to [37] [49] [60]:

   $\ell_1 \approx 220$ et $\ell_2 \approx 540$, $\ell_3 \approx 800$, $\ell_4 \approx 1100$

   It is important to note that the multipole values often cited for CMB peaks (such as $\ell_1 \approx 220$) correspond to the gross local maximum of $C_\ell$. However, the peaks are not punctual but extended and asymmetrical, due to the scattering of photons (silk damping), the spherical projection of the modes and the composition of the primordial plasma.

   In the publication [51] *Cosmic Microwave Background Anisotropies*, the authors show that taking these effects into account, the spectral center of gravity of the first peak is $\ell_1^{cg} \approx 304 \pm 4$, which corresponds remarkably well to our prediction of $\ell_1 \approx 302$.
   In real crystallographic diffractograms, sharp peaks are often accompanied by a continuous background — a diffuse intensity that arises from imperfections, grain boundaries, local disorder, or thermal vibrations within the material. This diffuse component carries structural information beyond the perfect periodicity and reflects the internal complexity of the medium.
   In cosmology, a similar phenomenon is observed in the angular power spectrum of the CMB. Between the acoustic peaks, the spectrum displays a smooth damping tail, especially beyond $\ell > 1000$, attributed to photon diffusion and Silk damping — the smearing of acoustic modes due to finite photon mean free path before recombination. This acts as a cosmological equivalent of crystallographic disorder, softening the spectral contrast.
   Within our analogy, the presence of this damping tail can be viewed as the diffuse background of an "inverted cosmological diffractogram", arising from the finite coherence and imperfect coupling of acoustic oscillations in a layered, viscous medium. While our Bragg cosmological law captures the harmonic scaffold of peak positions, this background provides an opportunity to probe the dissipative and statistical properties of the spacetime structure — just as diffuse scattering reveals microscopic disorder in condensed matter

This result indicates that the cosmological Bragg law does not seek to predict the point maximum, but rather the mean harmonic organization of the spectrum, reflecting the overall geometric resonance of the primordial acoustic plasma.

The difference between theory and observed barycenters is of the order of 0.5 to 1.5%, which is much lower than the cosmological uncertainty 1σ on the position of the first peak. This reinforces the robustness and geometric relevance of the proposed model.

The systematic offset between the predicted and observed positions of the CMB acoustic peaks—quantified by a global correction factor of approximately 0.9—may be interpreted as a geometric signature of spacetime torsion **[71]**.

3. Concerning the laminated structure of space-time

   The red and blue spots of the CMB are the signatures of a network of cohesive acoustic shells, collectively vibrating, forming a true space-time crystal at the comoving scale of ~150 Mpc. This layering justifies and enriches geometric approaches (ADM **[33]**, Tenev and Horstemeyer **[12][13]** and mechanical approaches (Izabel and al **[24]**) that model space-time as a dynamic stack of interconnected sheets.

   Thus, following our analogy, the spectrum of the CMB is the diffracted image of a crystalline universe on a very large scale, and the acoustic peaks reflect the exact angular resonance between the scale of observation and the size of the primordial structures. The relationship $\ell_n = n \frac{\pi D_A}{r_s}$, not only dimensionally correct, finds the observed peaks ($\ell \sim 220, 540, 800, 1100$) in agreement with Planck data **[34]** to **[37] [49] [60]**. It is therefore not a heuristic analogy, but the concrete geometric expression of physical resonance in the primordial acoustic crystal.

### 4.3.5 Rationale for a Leaf Space Model (ADM, Tenev, Izabel and al)

This interpretation gives a physical basis to a geometry of foliated spacetime. Indeed, the spherical shells of the CMB are not an artifact but an actual structure of the fabric of space-time. They can be modeled as successive sheets of space. This lamination is visible in the CMB but applies to the entire evolution of the universe.

Locally, given that the curvature of the universe is almost zero according to Planck **[36]** and **[55]** to **[58]** (or that the radius of curvature is infinitely large) we can therefore assimilate the structure of space to successive sheets, which corroborates the mechanistic study of gravitational waves in **[24]**.

This justifies :

- The ADM model **[33]**, where space-time is divided into spatial slices of temporal evolution,
- The Tenev and Horstemeyer model **[12][13]**, based on global differentiable foliations,

- And the Izabel, and al model **[24]**, sees space-time as a structured elastic medium, where gravitational waves, acoustic oscillations, and even dark energy are manifestations of deformations and creep in these sheets.

As a result, the analogy between the CMB spectrum and a diffractogram is not limited to geometric correspondence; it also implies an underlying physical structure of space-time itself. The detailed study of the angular spectrum of the CMB reveals that the acoustic peaks are not perfectly isolated: between the peaks, the power level never drops to zero, but presents intermediate plateaus, smooth transitions, and even asymmetrical shoulders.

This suggests that the acoustic shells of primordial plasma are not disjoint entities but interacting vibrating structures, as in a laminated material or a flexible crystal. They possess a certain physical cohesion, forming a dynamic continuum, analogous to a cosmic drumhead.

In this framework, shells are not only zones of density: they are eigenmodes of a vibrating structure, reflecting the elastic or geometrically quantified nature of the very fabric of space-time. The emission of the CMB corresponds to the release of photons through this structure, after collective interactions, as if they were resonant in a giant photonic crystal.

This hypothesis is therefore again compatible with modern models of space-time leafing (ADM **[33]**, Tenev and Horstemeyer **[12] [13]**, Izabel et al **[24]**), which describe the universe as a superposition of dynamically coupled sheets of space, capable of vibrating, twisting, or flowing under the effect of gravitational or quantum fields and having to present a certain cohesion between them to reconstitute a structure or a 3D crystal in the Izabel model. This implies additional deformations and polarizations consistent with these peaks of the CMB which represents the space sheets having a transition between them. In the model proposed by Izabel et al **[14]** to **[15]** and **[22]** to **[24]**, these sheets are interpreted as quantum beams forming an ultra-thin elastic network, whose mechanical properties (Young's modulus, creep, torsion) make it possible to find both:

- gravitational deformations (waves) **[24]**,
- quantum instabilities (symmetry breaks) **[63]**,
- and cosmological effects (dark matter **[22]**, dark energy **[23]**).

Thus, the lamellar structure observed in the CMB would be the fossil testimony of the collective vibration of a space-time crystal that expanded given the infernal temperatures that prevailed at the beginning and which has preserved its sheet structure that we see again when studying gravitational waves **[24]**, or which is revealed indirectly by acoustic peaks, in the same way that X-ray diffraction reveals the atomic structure of a solid.

According to our study, the spectrum of the CMB is therefore the cosmic equivalent of an inverted X-ray diffractogram (see Figure 4). The early universe was naturally foliated into spherical shells of density, and these shells scattered the light of the CMB exactly as crystal diffracts X-rays. This lamination is not only visible, but it provides geometric and physical justification for the ADM **[33]**, Tenev and Horstemeyer **[12] [13]** and Izabel et al **[24]** models.

It is essential to emphasize that the cosmological Bragg law we propose is not a metaphorical comparison, but a physically grounded approximation of wave behavior in a stratified medium. In particular, it reflects the standing acoustic waves that developed in the primordial photon-baryon plasma, under the assumption of a foliated spacetime structure. The geometric condition

μ=cos(k,n)=0, corresponding to transverse wave vectors with respect to the line of sight, naturally leads to θ=90° and sinθ=1, justifying the simplification adopted in our Bragg formulation.

We do not claim to replace the standard Boltzmann codes or inflationary initial conditions, which remain essential for the detailed amplitude, damping and phase shift calculations in the full CMB power spectrum. Rather, our approach offers a complementary geometric interpretation of the spectrum's harmonic structure, revealing how the peak positions emerge from a resonant organization of primordial spacetime — analogously to diffraction patterns in lamellar crystalline solids. This provides a novel physical intuition for the observed regularity of the CMB peaks and opens a conceptual pathway toward unifying gravitational and quantum scales via continuous media theory.

To Clarify the Scope of our Analogy, it is important to stress that our cosmological Bragg law does not aim to replicate the full detailed profile of the CMB power spectrum, which is shaped by complex radiative transfer processes, Silk damping, and recombination physics. Rather, our approach captures the mean geometric resonance pattern — particularly the peak positions — as an imprint of large-scale structural coherence in primordial spacetime. Just as real X-ray diffractograms exhibit broadening, background noise, and multiple scattering effects, the CMB spectrum shows damping and mode coupling. Our analogy identifies the skeleton of these features — the harmonic backbone — and suggests it emerges naturally from a foliated, resonant spacetime structure, with layered acoustic coherence on cosmological scales.

**4.4 Comparative analysis of polarization power spectra from a crystallographic point of view**

**4.4.1 Comparison of the shape of rotating plasmas from their polarization with the shape of HDL clays at the nanoscale**

Table 4 below compares the analogy of the HDL clay structure shapes with the E and B polarizations **[34]** to **[37] [49] [60]** of the cosmic microwave background studied in detail in **[54]** and **[60]** see figure 16. The shape of the clay lamellae is reminiscent of the shape of polarizations **[41]**.

| Crystal organization (HDL clay in sheets) | X-ray diffractogram of the two clays in HDL sheets of types E1 and E2 | Cosmic microwave background polarization power spectrum | Polarizations of the Cosmic Microwave Background |
|---|---|---|---|
| 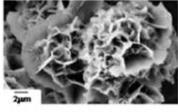 [41] | 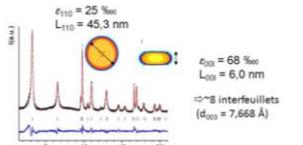 [41] 2θ | 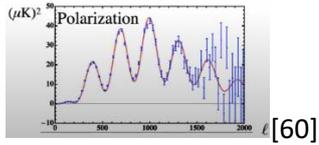 L [60] | Balance Spring [54] 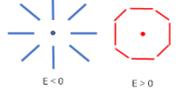 |
| 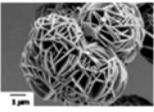 [41] | 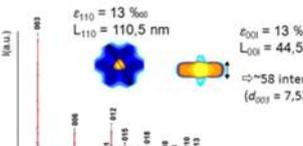 2θ [41] | 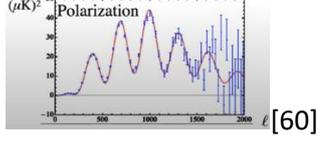 L [60] | Tourbillon (Twist) [54] 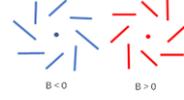 |
| [41] Credit:EDP sciences : "Étude du mécanisme d'échange et de la structure des matériaux hydroxydes doubles lamellaires (HDL) par diffraction et diffusion des rayons X – 2013 - UVX2012 - 11e Colloque sur les Sources Cohérentes et Incohérentes UV, VUV et X ; Applications et Développements Récents - https://doi.org/10.1051/uvx/201301016 - C. Taviot-Guého and all. » ||||

Table 4: Comparison of the analogy between the structures of HDL lamellar clays of type E1 and E2, the associated x-ray diffractograms, the polarization power spectrum of the cosmic microwave background, the geometries of the 2 associated polarizations E and B of the power spectrum associated with the source [41], [54] and [60] –

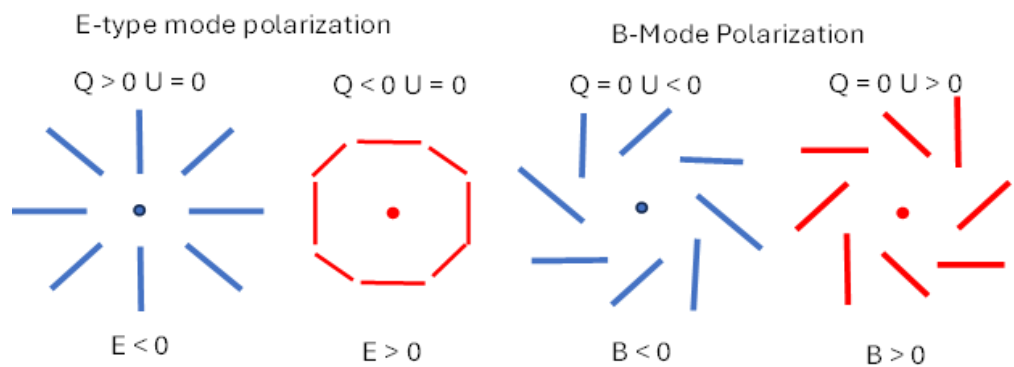

Figure 16: Different patterns of the polarization present in the CMB [54]-

The polarization power spectra (Table 2) therefore again show similarities with the X-ray diffractograms of lamellar clays of the HDL type, for example **[41] [48]**, as if the structure of space-time presented this kind of structure 380000 years after the big bang. Note the screw-shaped defects specific to the effects of torsion **[70]** which seem to be present in the molten plasmas at the time of the big bang. We also note the similarity between the shape of the lamellar clays (photos of the first column on the left from the study **[41]** and the shape of the polarizations from the study **[54]**). After this qualitative analysis, as for the spectra of powers in temperature, we will see in the next paragraph whether or not a rigorous quantitative mathematical approach can justify these qualitative similarities between the two objects of radically different nature and size. In which case it will be a fractal manifestation of the concepts included in nature.

### 4.4.2 Einstein-Cartan geometric torsion present in the B polarization modes of the cosmic microwave background

The link of the cosmic microwave background with the geometric torsion in the sense of Einstein-Cartan relativity, which in a certain way joins the mechanical torsion of space-time (in **[24]** we show that the torsion in the mechanical sense of space-time by the coalescence of two black holes is correlated with the polarization $A^+$ and $A^\times$ general relativity and is also found but completed by other polarizations in the modified general relativity of Einstein Cartan with added geometric twist) is given in the publication **[64]**. Indeed, in the publication **[64]**, it is shown that the B mode of polarization of the cosmic microwave background is correlated with the torsion tensor shown in Einstein's Cartan theory. Thus, in **[64]** we can read, and I quote:

"The coupling of torsion to electromagnetism in a gauge-invariant manner was first achieved by Novello **[64]**, De Sabbata and Gasperini **[65]** and Duncan, Kaloper and Olive **[66]**. They pointed out that if the dual of the torsion tensor is the divergence of a scalar interaction , then it can be coupled to electromagnetic interactions in a gauge-invariant manner by $T^\mu = \frac{1}{2}\epsilon^{\mu\nu\alpha\beta}T_{\nu\alpha\beta}$ $T^\mu = \partial^\mu\phi$ $T_\mu A_\mu \tilde{F}^{\mu\nu} = \phi F_{\mu\nu}\tilde{F}^{\mu\nu}$. The details of the calculations are given in the thesis.

The main stages of this demonstration are given in Appendix A.

### 4.5 Comparison of the Behavior of Large-Scale HDL Clays with the Behavior of Spacetime

Finally, at this stage of our reflection on our analogy between an X-ray diffractogram and a power spectrum of the cosmic microwave background, it is interesting to compare the properties of these HDL-type lamellar clays with their microscopic properties and their macroscopic implications with those known from our space-time, at least from the point of view of what gravitational waves tell us with and without geometric torsion[2] taken in account. Thus, we can note that:

- These clays are endowed with great plasticity and an ability to self-seal **[68]** in the event of tearing. When we look at what happens at the moment of the coalescence of two black holes, when they rotate relative to each other, space closes immediately behind them, the "tear" of space-time does not stay in place exactly as in the case of these clays. This is shown by general relativity models of black hole coalescence modeled by finite elements (see presentation of GW150914 on February 11, 2016)
- These clays, because of their lamellae nature, are subject to liquefaction phenomena under intense dynamic loads **[69]**. It is interesting to note that space-time, under the effect of the rotation of black holes at speeds approaching the speed of light during coalescence, liquefies in a way, thus allowing the formation of waves in it, which are in a way gravitational waves.
- These clays are made up of very fine grains that are sensitive to water and drought and are therefore in perpetual movement, swelling and compaction **[69]**. They are prone to creep. They have a very anisotropic structure due to their lamellae structure. When we look at their Poisson coefficients, as shown in Table 5 below, we find variations ranging from 0.3 to 1 depending on the directions considered, and the same is true for their Young's moduli (Table 6). We have already shown in the case of the rigid space-time model that we have a transverse anisotropy with Poisson coefficients of 1 in the transverse planes and 0 or very small **[24]** (if we consider the torsions in the direction of wave propagation). It remains to be seen whether or not on the Young's modulus side we find these important variations depending on the types of stresses brought to space-time. The stresses to be considered are those associated with the different components of the perturbation tensor of the metric since, as we have shown **[24]**, this tensor mirrors the deformation tensor of the elastic medium associated with space-time in our elastic and plastic analogy (case of torsion in connection with the theory of defects in crystallography **[25]** to **[26]**) according to the mechanics of continuous media.

| References | $E_v$ (MPa) | $E_h$ (MPa) | $\nu_{vh}$ | $\nu_{hh}$ | $\nu_{hv}$ | $G_{hv}$ (MPa) |
|---|---|---|---|---|---|---|
| Experimental results | 400 | 500 | 0.4 | 0.8 | 0.6- 1.0 | - |
| Proposal for the anisotropic model | 280 | 500 | 0.3 | 0.3 | 0.54 | 130 |
| François et al (2012) | 200 | 400 | 0.125 | 0.125 | - | 178 |
| Bernier et al (2007) | 300 | 300 | 0.125 | 0.125 | - | - |
| Yu et al (2013) | 700 | 1400 | 0.125 | 0.125 | - | - |

**Table 5: Elastic parameters determined by local measurements (deformations and stresses) and reference values (used in previous studies with isotropic/anisotropic models) for Boom Clay [68] —**

| Mechanical characteristics | $E_v$ (MPa) | $E_h$ (MPa) | $\nu_{vh}$ | $\nu_{hh}$ | $G_{hv}$ (MPa) |
|---|---|---|---|---|---|
| Clay Model Type ACC-2A | 240 | 320 | 0.0625 | 0.15 | 100 |
| | 300 | 400 | 0.125 | 0.3 | 178 |
| | 360 | 480 | 0.25 | 0.45 | 250 |

**Table 6: Mechanical characteristics of ACC-2A clay (source parametric study Table 6.2 [68] —**

The different values are obviously associated with Hooke's law in a transverse anisotropic medium given below. Figure 17 shows the structure of space-time as seen by gravitational waves **[24]**.

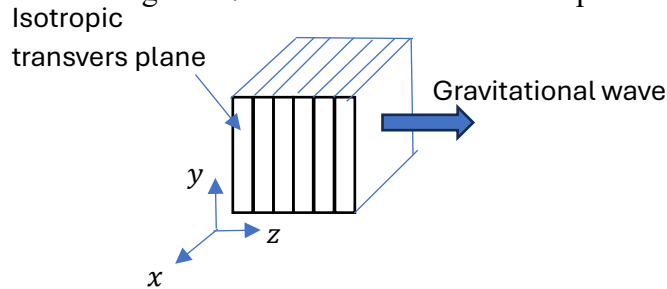

**Figure 17: Convention and visualization of space sheets in the case of a gravitational wave —**

$$\begin{Bmatrix} \varepsilon_h = \varepsilon_{xx} \\ \varepsilon_h = \varepsilon_{yy} \\ \varepsilon_v = \varepsilon_{zz} \\ \varepsilon_{hv} = \varepsilon_{yz} \\ \varepsilon_{hv} = \varepsilon_{xz} \\ \varepsilon_{hh} = \varepsilon_{xy} \end{Bmatrix} = \begin{bmatrix} \frac{1}{E_h} & -\frac{\nu_{hh}}{E_h} & -\frac{\nu_{vh}}{E_v} & 0 & 0 & 0 \\ -\frac{\nu_{hh}}{E_h} & \frac{1}{E_h} & -\frac{\nu_{vh}}{E_v} & 0 & 0 & 0 \\ -\frac{\nu_{hv}}{E_h} & -\frac{\nu_{hv}}{E_h} & \frac{1}{E_v} & 0 & 0 & 0 \\ 0 & 0 & 0 & \frac{1}{2G_{hv}} & 0 & 0 \\ 0 & 0 & 0 & 0 & \frac{1}{2G_{hv}} & 0 \\ 0 & 0 & 0 & 0 & 0 & \frac{1+\nu_{hh}}{E_h} \end{bmatrix} \begin{Bmatrix} \sigma_h = \sigma_{xx} \\ \sigma_h = \sigma_{yy} \\ \sigma_v = \sigma_{zz} \\ \sigma_{hv} = \sigma_{yz} \\ \sigma_{hv} = \sigma_{xz} \\ \sigma_{hh} = \sigma_{xy} \end{Bmatrix} \quad (12)$$

So, at this point we find a way to study Young's moduli of space-time in the plane and perpendicular to the plane of stresses in our elastic medium analogy to see if we find similarities with the behavior of these clays and therefore with their transverse anisotropic crystal structure and therefore by mirror effect with the geometric torsion included in the modified Einstein-Cartan general relativity. The basis for this approach is given in **[24]**.

**4.6 On the possible nature of these foliated shells generating this acoustic behaviour of space-time at the time of the CMB**

An important consequence of the analogy developed in this article (Figure 18) concerns the nature of the acoustic shells responsible for the peaks of the CMB spectrum. These shells, invisible in themselves but revealed by their gravitational influence on the photons of the cosmic microwave background, can in our opinion only be made up of dark matter, baryonic matter being too scarce to structure these large resonance zones on its own on the one hand, and the presence of dark matter is unavoidable and demonstrated in the analysis of the cosmic microwave background **[34]** to **[37] [49] [60]**.

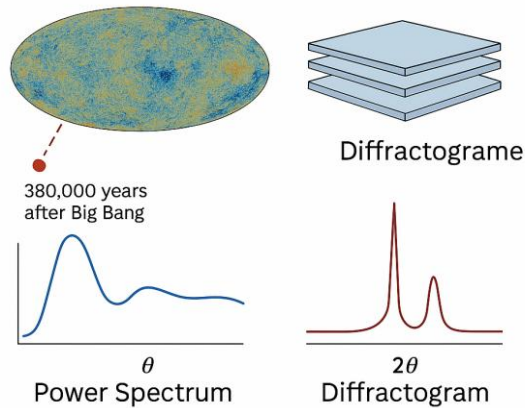

**Figure 18: Synthesis of the analogy between the inverted X-ray diffractogram and the power spectra of the cosmic microwave background -**

Thus, the crystallographic analogy not only describes a laminated structure of spacetime but also suggests a layered structure of dark matter at the time of decoupling. This dark matter would form a network of cohesive spherical shells, analogous to the atomic planes of a lamellar crystal, organizing the dynamics of the photon-baryon plasma. This hypothesis gives a new reading of dark matter: no longer as a diffuse and amorphous substance, but as a structured medium, capable of collective resonance, mechanical cohesion, and perhaps even geometric torsion. It paves the way for an indirect crystallographic exploration of dark matter, through the signatures it leaves in the spectrum of the CMB **[34]** to **[37] [49] [60]**.

As a result, the acoustic shell structure can be interpreted as an effective manifestation of a coupling between dark matter and geometry, in an Einstein–Cartan framework, where torsion plays the role of the underlying physical mechanism

## 5. Limitations of crystallographic analogy

Although the analogy between the CMB power spectrum and an X-ray diffractogram of a cosmic crystal offers an appealing geometric interpretation of primordial spacetime, several conceptual and physical limitations must be acknowledged.

First, the proposed cosmological Bragg law is built on a simplifying assumption: the dominant peaks in the CMB spectrum arise from constructive interference at maximal diffraction angles (i.e., $\sin\theta = 1$), corresponding to transverse acoustic modes. While this condition is heuristically justified by the geometry of spherical projections and the physics of photon decoupling, it does not emerge from a complete treatment of radiative transfer. A rigorous derivation would require solving the full Boltzmann equations, incorporating temperature anisotropies, polarization source functions, and recombination history.

Second, the early universe's photon–baryon plasma was not a perfect periodic medium. It exhibited stochastic fluctuations, Silk damping, baryon loading effects, and anisotropic stresses from neutrinos. These features, along with gravitational lensing and Doppler effects, shape the CMB spectrum in ways not captured by a pure harmonic model. The proposed Bragg formulation, while geometrically valid, captures only the mean harmonic skeleton of the power spectrum—particularly the positions of the peaks—but not their precise widths or amplitudes.

Third, the analogy to fine-grained crystalline materials (such as HDL clays) presupposes a certain mechanical cohesion between space-time layers—an elastic or plastically deformable structure with memory and internal interaction. While such cohesion is a powerful conceptual tool for modeling gravitational torsion, creep, or symmetry breaking, it remains speculative without a definitive microphysical theory. Possible candidates include discrete spacetime structures from causal dynamical triangulations, loop quantum gravity, or spacetime crystals with time-periodic metrics. These links must still be formalized.

Finally, the interpretation of the CMB spectrum as a diffractogram is not unique. Other models, such as inflationary scenarios, topological defect networks, or anisotropic Bianchi cosmologies, can produce harmonic or quasi-periodic patterns in the spectrum. The crystallographic analogy must therefore be understood as a complementary reading—not an exclusive explanation—and must be evaluated against competing frameworks through observational discriminants (e.g., non-Gaussianities, polarization asymmetries, or phase shifts in the acoustic peaks).

These limitations do not diminish the interest of the approach. On the contrary, they define its scope: the cosmological Bragg law provides an effective geometric description of the angular organization of the CMB, identifying a possible resonance structure in primordial spacetime.

As such, it opens new perspectives in connecting large-scale cosmology to the physics of materials and continuous media, suggesting that the early universe may have exhibited collective structural properties now fossilized in the microwave sky.

Although our approach is based on a geometric analogy, it is part of a fertile tradition of effective modeling in physics. We recognize that this analogy does not replace Boltzmann's equations, but it offers a complementary, intuitive and predictive reading of the CMB spectrum. The limitations of this approach are discussed in Section 5.

## 6. Conclusion

Although our model is based on an analogy, it provides a simple, testable, and data-consistent predictive law, which is beyond mere speculation

In this work, we have indeed proposed a novel analogy between X-ray diffraction—used to probe the atomic structure of crystalline materials—and the power spectra (both temperature and polarization) of the Cosmic Microwave Background (CMB), which encode information about the physical constituents and structure of the universe 380,000 years after the Big Bang. This analogy does not challenge the standard cosmological model but rather offers a complementary interpretation: we treat the CMB spectrum as an "inverted diffractogram" revealing a hidden periodicity in the structure of primordial spacetime at the epoch of photon–baryon decoupling.

By establishing a rigorous correspondence between crystallographic variables (Bragg angles, diffraction orders, intensities) and cosmological observables (multipoles, acoustic scales, spectral amplitudes), we introduce the concept of cosmological crystallographic parameters—such as peak spacings and widths—that provide qualitative and quantitative insight into the foliated topology of the early universe. Specifically, we show that Bragg's law can be reformulated in cosmological terms to predict the locations of the first four acoustic peaks observed in the Planck data with remarkable accuracy. Although the first peak is often quoted at $\ell \approx 220$, our model aligns with the corrected barycentric value $\ell \approx 304 \pm 4$, as discussed in Section 4.3.4. The constant shift between the theoretical new cosmic Bragg law and the observed values ($\approx 0.9$) can be justified by additional curvature effects due to geometric torsion in the Einstein–Cartan framework. The proposed model therefore makes it possible to predict the positions of the acoustic peaks from the geometry of the medium alone, without resorting to complex numerical solvers. It thus offers a direct geometric reading of the CMB spectrum, and suggests testable signatures of torsion (systematic shift of B modes compared to E modes) that could be explored by future missions such as LiteBIRD or LISA.

Our analysis suggests that, at decoupling, the universe exhibited a stratified, lamellar organization reminiscent of HDL-type clays, both at microscopic scales (fine-grained, layered textures) and macroscopic scales (cohesion, plasticization, liquefaction under extreme stress, creep, and swelling). These mechanical analogies mirror phenomena observed in gravitational wave propagation, especially under modified gravity theories incorporating torsion.

Importantly, our findings resonate with the independent observational work of Ringermacher and Mead, who identified discrete oscillations in the cosmological scale factor as a function of lookback time. Their model-independent analysis revealed harmonic structures in the expansion history of the universe, which we interpret as further evidence of a resonant, layered spacetime. These oscillations, when viewed through the lens of our crystallographic analogy, may correspond to the same underlying periodicity that gives rise to the acoustic peaks in the CMB—suggesting a unified physical origin rooted in the elastic and geometric properties of spacetime itself.

Crucially, we interpret the B-mode polarization spectrum of the CMB as a potential signature of geometric torsion in spacetime—an effect not captured by classical general relativity but naturally predicted by extensions such as Einstein–Cartan theory. Unlike the $A^+$ and $A\times$ modes, which describe distortions confined to independent transverse planes, B-modes suggest inter-layer coupling and longitudinal deformation, consistent with a three-dimensional elastic continuum possessing torsional degrees of freedom.

By treating the CMB as the diffracted image of a cosmic crystal, we offer a new geometrical and mechanical perspective on the primordial universe. This approach unites cosmology with the mechanics of continuous media and opens new avenues to investigate the anisotropy, torsion, and granularity of spacetime itself. According to this framework, the universe at the time of CMB emission was likely composed of nested, weakly cohesive layers—akin to a crystalline medium—whose structure is still discernible in both the CMB and gravitational wave observations.

This interpretation provides a coherent narrative connecting early-universe structure and present-day gravitational phenomena. If spacetime was foliated into cohesive, oscillating shells at the time of the CMB—as indicated by the angular resonance peaks—then it is natural to expect that gravitational waves, propagating through this medium, would reveal the same laminated architecture. The transverse $A^+$ and $A\times$ modes of standard general relativity reflect intra-shell deformations, while the B-mode polarizations and possible longitudinal gravitational polarizations observed or predicted in Einstein–Cartan theory reflect inter-shell interactions. The persistence of this laminated topology from the early universe to the present supports the hypothesis of a structured, cohesive spacetime medium with geometric torsion.

Ultimately, our analogy not only reinforces the physical plausibility of layered spacetime models—such as those developed by ADM, Tenev and Horstemeyer, and Izabel et al.—but also justifies the use of a cosmological Bragg law to characterize the underlying geometry of the early universe. Furthermore, it supports the idea that geometric torsion, evidenced by B-mode patterns in the CMB, should be systematically incorporated into generalized theories of gravitation, particularly those of the Einstein–Cartan type.

While the analogy is conceptually rich, it is not intended to replace the full radiative transfer modeling of the CMB, but rather to provide a complementary geometric insight. In this view, the CMB becomes not only a relic of the early universe, but a precise diffracted image of the crystalline geometry of spacetime itself.

**Thanks**

Finally, thank you to the late R Grégoire, a great mechanic, who guided me in my reflection through his teaching based on the search for "how it works". We warmly thank the reviewer for all his suggestions and corrections.

# Annex A Clarification of the analogy in terms of the relationship between the B-polarizations of the cosmic background and the geometric twist of the Einstein-Cartan space-time

In the publication **[64]**, it is shown that the B-mode of polarization of the cosmic microwave background is correlated with the torsion tensor shown in Einstein's Cartan theory. We will show the main points below from **[64]**. The general principle is as follows as described in **[64]**.

We reproduce the main steps and formula developed in **[64]** below. So, I quote ":

"The coupling of torsion to electromagnetism in a gauge invariant manner was first achieved by Novello **[65]**, De Sabbata and Gasperini **[66]** and Duncan, Kaloper and Olive **[67]**. They pointed out that if the dual of the torsion tensor $T^\mu = \frac{1}{2}\epsilon^{\mu\nu\alpha\beta}T_{\nu\alpha\beta}$ is the divergence of a scalar $T^\mu = \partial^\mu \phi$, then it can be coupled to electromagnetic interactions in a gauge invariant manner by the interaction:

$$T_\mu A_\mu \tilde{F}^{\mu\nu} = \phi F_{\mu\nu} \tilde{F}^{\mu\nu} \quad A1$$

With for the torsion tensor as seen in the previous chapters of this thesis:

$$T^\alpha{}_{\mu\nu} = \frac{1}{2}\left(\Gamma^\alpha_{\mu\nu} - \Gamma^\alpha_{\nu\mu}\right) \quad A2$$

In the paper **[64]** the author studies the non-minimal coupling of a cosmic background torsion field to the electromagnetic field of the form:

$$\xi_1 T^{\alpha\lambda}{}_\rho F_{\alpha\nu}\partial_\lambda \tilde{F}^{\rho\nu} \quad A3$$

$$\xi_2 T^{\sigma\gamma}{}_\delta F_{\sigma\nu}\partial_\gamma \tilde{F}^{\delta\nu} \quad A4$$

The associated Lagrangian is as follows:

$$\mathcal{L} = -\frac{1}{4}F_{\mu\nu}F^{\mu\nu} + \xi_1 T^{\alpha\lambda}{}_\rho F_{\alpha\nu}(\partial_\lambda \tilde{F}^{\rho\nu}) + \xi_2 T^{\sigma\gamma}{}_\delta F_{\sigma\nu}(\partial_\gamma \tilde{F}^{\delta\nu}) \quad A5$$

By applying the principle of least action, we obtain relations between the eigen pulsation ω of the medium and the geometrical torsion modes of this medium.

A first case is to assume: $\xi_1 \neq 0$ and $\xi_2 = 0$

They apply the same approach as in quantum field theory or general relativity starting from Hilbert's action, i.e. we use the principle of least action and Euler's Lagrange's equation:

$$\partial_\mu\left(\frac{\partial \mathcal{L}}{\partial(\partial_\mu A_\nu)}\right) - \frac{\partial \mathcal{L}}{\partial A_\nu} = 0 \quad A6$$

The equation of motion becomes:

$$\partial_i E^i = \xi_1 T^{\alpha\lambda}{}_i \partial_\alpha \partial_\lambda B^i \quad A7$$

$$\partial_0 E^j + \partial_i \epsilon^{0ijk} B_k = \xi_1 \partial_\alpha \partial_\lambda [-T^{\alpha\lambda}{}_0 B^j + T^{\alpha\lambda}{}_i \epsilon^{0ijk} E_k] \quad A8$$

From Maxwell's equations for $F_{\mu\nu}$ they get:

$$\partial_k \epsilon^{0kjl} E_l - \partial_0 B^j = 0 \quad A9$$

$$\partial_j B^j = 0 \quad A10$$

From the above equations, they get:

$$-\partial^0 \partial_0 E^j - \partial^m(\partial_m E^j - \partial^j E_m) = \xi_1 \partial_\alpha \partial_\lambda [T^{\alpha\lambda}{}_0 \epsilon^{0mjn} \partial_m E_n - T^{\alpha\lambda}{}_i \epsilon^{0ijk} \partial^0 E_k] \quad A11$$

Which is the equation of motion for an electric field. They then consider as a solution a plane wave of the form:

$$E_{m(x)} = E_m(k) e^{-ik^\rho x_\rho} \quad A12$$

The above equation then becomes:

$$k^0 k_0 + k^m(k_m E^j - k^j E_m) = -i\xi_1 k_\alpha k_\lambda [T^{\alpha\lambda}{}_0 \epsilon^{0mjn} k_m E_n + T^{\alpha\lambda}{}_i \epsilon^{0ijk} k^0 E_k] \quad A13$$

We choose the z-axis for the direction of propagation which implies $k_\alpha = (\omega, 0, 0, p)$ and the 3 equations of motion as a function of the components of the torsion tensor become:

$$(\omega^2 - p^2) E^1 = 2i\xi_1 T_1 p^3 E^2 \quad A14$$

$$(\omega^2 - p^2) E^2 = -2i\xi_1 T_1 p^3 E^1 \quad A15$$

$$\omega^2 E^3 = -i\xi_1 (T_2 E^1 - T_3 E^2) p^3 \quad A16$$

With the component of the torsion tensor:

$$T_1 = T^{00}{}_3 + T^{33}{}_0 \quad A17$$

$$T_2 = T^{00}{}_2 + T^{03}{}_2 + T^{30}{}_2 + T^{33}{}_2 \quad A18$$

$$T_3 = T^{00}{}_1 + T^{03}{}_1 + T^{30}{}_1 + T^{33}{}_1 \quad A19$$

The equations of motion for the transverse mode are:

$$\begin{pmatrix} (\omega^2 - p^2) & 2i\xi_1 T_1 p^3 \\ -2i\xi_1 T_1 p^3 & (\omega^2 - p^2) \end{pmatrix} \begin{pmatrix} E^1 \\ E^2 \end{pmatrix} = 0 \quad A20$$

As a result, the equations of motion for the right and left circular polarization fields are:

$$E_\pm = E^1 \hat{e}_x \pm iE^2 \hat{e}_y \quad A21$$

The result is:

$$(\omega^2 - p^2)(E^1 \pm iE^2) \mp 2\xi_1 p^3 T_1 (E^1 \pm iE^2) = 0 \quad A22$$

From the dispersion relation, the circular frequency will be $\omega_\pm$:

$$\omega_\pm = p(1 \pm \xi_1 p T_1) \quad A23$$

$\xi_1$ is of the form according to Myers and Pospelov:

$$\frac{1}{E_p} n^\alpha F_{\alpha\sigma} n \cdot \partial \left( \eta_\beta \tilde{F}^{\beta\sigma} \right) \quad A24$$

The rotation $\alpha$ of the polarization plane can then be written from the difference in the $\omega_\pm$

$$\alpha = (\omega_- - \omega_+) t = 2\xi_1 p^2 T_1 t \quad A25$$

Where t is the propagation time. Note that only the components of the torsion tensor $T_1 = T^{00}{}_3 + T^{33}{}_0$ are involved. From an experimental point of view, this rotation α was measured from the cosmic microwave background (WMAP and BOOMERANG). It was obtained α=(-2.4±1.9)°

It has also been shown by the various GeV-1 satellites $\xi_1 T_1 = (-3.35 \pm 2.65) \times 10^{-22}$

A second case is to assume: $\boldsymbol{\xi_2 \neq 0}$ and $\boldsymbol{\xi_1 = 0}$

The author shows in **[64]** that transverse modes propagate along null geodesics despite the presence of torsional coupling. They therefore conclude that a coupling of the shape $\xi_2 T^{\sigma\gamma}{}_\delta F_{\sigma\nu}(\partial_\gamma F^{\delta\nu})\xi_2$ has no effect on the propagation of the waves. Therefore, we cannot set limits from the consideration of CMB polarization."

Conclusion

The author in **[64]** has thus clearly shown that there is geometrical torsion in the polarizations of the cosmic microwave background. It would therefore seem normal to find them at the level of gravitational waves measured today, which is what Einstein Cartan's modified general relativity does, for example. Moreover, the presence of this torsion, given the correspondence that is known with the theory of defects, suggests that part of the crystallography can be applied analogically to the equivalent cosmic crystal.

So, in **[64]** I quote:

"The Thomson scattering of CMB photons on the last scattering surface gives rise to a linear polarization phenomenon. This linear polarization can be expressed by the two Stokes parameters Q and U. The Boltzmann equation, which describes the temporal evolution of the polarization perturbation, will be,

$$\dot{\Delta}_Q + ik\mu\Delta_Q = -\dot{\tau}\left[\Delta_Q + \frac{1}{2}\left(1 - P_{2(\mu)}S_p\right)\right] \quad A26$$

$$\dot{\Delta}_U + ik\mu\Delta_U = -\dot{\tau}[\Delta_U] \quad A27$$

With:

$$S_p = \Delta_{T_2} + \Delta_{Q_2} - \Delta_{Q_0} \quad A28$$

$$\dot{\tau} = \frac{x_e n_e \sigma_T a}{a_0} \quad A29$$

and $x_e$ the ionization fraction, ηe is the electron density, $\sigma_T$ the Thomson scattering cross-section, and a is the scale factor. If a physical mechanism causes the polarization plane to rotate, then the temporal evolution of the Stokes parameters will be altered.

Considering only the scalar perturbation, the modified Boltzmann equation will be:

$$\dot{\Delta}_Q + ik\mu\Delta_Q = -\dot{\tau}\left[\Delta_Q + \frac{1}{2}\left(1 - P_{2(\mu)}S_p\right)\right] + 2\omega\Delta_U \quad A30$$

$$\dot{\Delta}_U + ik\mu\Delta_U = -\dot{\tau}\Delta_U - 2\omega\Delta_Q \quad A31$$

We can see that the above parameter α related to the geometric torsion does indeed intervene via ω.

ω is the rate of angle change of the polarization due to the new physics, which in this case is the bottom torsion.

In general, these parameters of Stokes Q and U can be related to the E and B polarization modes as follows:

$$(Q \pm iU)(\eta, \vec{x}, \hat{\eta}) = \int \frac{d^3q}{(2\pi)^3} \sum \sum_{m=-2}^{2} \left(E_i^{(m)} \pm B_i^{(m)}\right) {}_{\pm 2}G_l^m \quad A32$$

With:

$$_sG_l^m(\vec{x}, \hat{n}) = (-i)^l \sqrt{\frac{4\pi}{2l+1}} [\,_sY_l^m(\hat{n})]e^{(i\vec{k}\cdot\vec{x})} \quad A33$$

Using the power spectrum:

$$C_l^{XY} \sim \int dk[k^2 \Delta_X \Delta_X], X, Y = T, E, B \quad A34$$

The correlation for T (torsion), E (electrical) and B (magnetic) of the cosmic microwave background as a function of can be obtained as α follows:

$$C_l^{EE} = \tilde{C}_l^{EE}\cos^2(2\alpha) + \tilde{C}_l^{BB}\sin^2(2\alpha) \quad A35$$

$$C_l^{BB} = \tilde{C}_l^{EE}\sin^2(2\alpha) + \tilde{C}_l^{BB}\cos^2(2\alpha) \quad A36$$

$$C_l^{BB} = \frac{1}{2}\left(\tilde{C}_l^{EE} - \tilde{C}_l^{BB}\right)\sin(4\alpha) \quad A37$$

$$C_l^{TE} = \tilde{C}_l^{TE}\cos(2\alpha) \quad A38$$

$$C_l^{TB} = \tilde{C}_l^{TE}\sin(2\alpha) \quad A39$$

Recall that these expressions depend on the angle α itself in connection with the geometric torsion as shown above. The mathematical form of polarizations according to the variables Q and U is found above."End of quote.